\documentclass[aps,prb,twocolumn,amsmath,amssymb,nofootinbib,superscriptaddress,floatfix]{revtex4-1}
\def\beq{\begin{equation}}
\def\eeq{\end{equation}}
\usepackage{amssymb}
\usepackage[hypertex]{hyperref}

\usepackage{color}
\usepackage{graphicx}

\begin{document}

%\title{Electromagnetic and gravitational responses of topological insulators 
%and superconductors in three dimensions}

\title{Electromagnetic and gravitational responses and anomalies 
in
topological insulators 
and superconductors}

\author{Shinsei Ryu}
\affiliation{
Department of Physics, University of California, Berkeley, California 94720, USA
}

\author{Joel E.\ Moore}
\affiliation{
Department of Physics, University of California, Berkeley, California 94720, USA
}
\affiliation{Materials Sciences Division, Lawrence Berkeley National Laboratory, Berkeley, CA 94720, USA}

\author{Andreas W.\ W.\ Ludwig}
\affiliation{
Department of Physics, University of California, Santa Barbara, California 93106, USA
}

\date{\today}

\begin{abstract}
One of the defining properties of the conventional three-dimensional 
(``$\mathbb{Z}_2$-'', or ``spin-orbit''-) topological insulator
is its characteristic magnetoelectric effect, as described by axion electrodynamics. 
In this paper, we discuss an analogue of such a magnetoelectric 
effect in the thermal (or gravitational) and the magnetic dipole 
responses in all symmetry classes which admit
topologically non-trivial insulators or superconductors to exist
in three dimensions. 
In particular, for topological superconductors (or superfluids)
with time-reversal symmetry which lack $SU(2)$
spin rotation symmetry (e.g. due to spin-orbit interactions),
such as the B phase of $^3$He, 
the thermal response is the only probe which can detect 
the non-trivial topological character through transport. 
We show that, for such topological superconductors, 
applying a temperature gradient produces a thermal- (or mass-) 
surface current perpendicular to the thermal gradient. 
Such charge, thermal, or magnetic dipole responses provide 
a definition of topological insulators and superconductors
beyond the single-particle picture. 
Moreover we find, for a significant part
of the `ten-fold' list of topological insulators 
found in previous work in the absence of interactions,
that in general dimensions the effective field theory
describing the space-time responses is governed
by a field theory anomaly.
Since anomalies are known to be insensitive to whether the underlying fermions are interacting or not,
this shows that the classification of these topological insulators
is robust to adiabatic deformations by
interparticle interactions in general dimensionality.
In particular, this applies to symmetry classes DIII, CI, and AIII 
in three spatial dimensions, 
and to  symmetry classes D and C in two spatial dimensions.
\end{abstract}


\maketitle

\section{Introduction} 
\label{Label-Introduction}

The considerable recent progress in understanding topological insulating phases in three dimensions was 
initiated by studies of single-particle Hamiltonians describing electrons with time-reversal invariance.  
\cite{kanemele2-2005,fukanemele-2007,fu&kane2-2006,essinmoore,moore&balents-2006}
In both, two and three dimensions, time-reversal invariant Fermi systems 
which have topological invariants of $\mathbb{Z}_2$ type
are known to exists:insulators can be classified as ``ordinary'' or ``topological'' 
by band-structure integrals similar to the integer-valued integrals that appear 
in the integer quantum Hall effect.  
\cite{tknn,kohmoto} 
These invariants survive when disorder is added to the system. 
In fact, stability to disorder is one of defining properties of 
topological insulating phases (and also topological superconductors). 
The complete classification of topological insulators and topological superconductors
in any dimension has been obtained in Refs.\ \onlinecite{SRFL} and \onlinecite{kitaevclassification}, 
and in every dimension five of the ten Altland-Zirnbauer symmetry classes 
\cite{Zirnbauer96,Altland97} 
of single-particle Hamiltonians (including some describing the Bogoliubov quasiparticles of superconductors or superfluids, rather than ordinary electrons) contain topological insulating phases with 
topologically protected gapless surface states.

An important question is how these various phases can be defined in terms of a physical response function.  
Aside from aiding in experimental detection, such definitions also indicate that the phase is well-defined 
in the presence of interactions.  
The best studied example is the conventional three-dimensional 
(``$\mathbb{Z}_2$-'', or ``spin-orbit''-)
topological insulator with no symmetries beyond time-reversal, 
which has been recently observed in various materials including 
Bi$_x$Sb$_{1-x}$ alloys\cite{hsieh}, 
Bi$_2$Se$_3$, and Bi$_2$Te$_3$.  
\cite{hsieh09,Xia09,Hsieh09b,Chen09}
Such materials support a quantized magnetoelectric response generated by the orbital motion of the electrons, 
i.e., the phase can be defined by the 
response of the bulk polarization
to an applied magnetic field.
\cite{qilong,essinmoorevanderbilt}
The possibility of 
such a bulk response 
was discussed some time ago as a condensed matter realization of ``axion electrodynamics''.
\cite{wilczekaxion}

The first goal of this paper is 
to find, for all three-dimensional topological insulators and superconductors,
the corresponding responses
 that result from the coupling of the theory to gauge and 
gravitational
\cite{footnote0}
fields.
The second goal of this paper is to understand to what extent the classification
scheme found previously for topological insulators of non-interacting fermions
can be stable to fermion interactions. This addresses the question
as to whether certain topological insulators which describe distinct topological
phases in the absence of fermion interactions (connected only by quantum 
phase transitions at which the bulk gap closes), 
can be adiabatically deformed into each other
when interactions are included 
(without closing the bulk gap). 
We find that this cannot happen
e.g. in symmetry classes DIII, CI and AIII in three spatial dimensions,
and in symmetry classes D and C in two dimensions.
More generally, 
in the last (more technical) chapter of this paper we provide 
an answer to this question in general dimensionalities 
for a significant part of the list of topological insulators
(superconductors) within the `ten-fold' classification scheme, obtained
for non-interacting particles\cite{SRFL,SRFLnewJphys,SRFLLandau100,kitaevclassification}. 
In particular, we relate the topological features of these topological insulators 
to the appearance of a topological term in the effective field theory 
describing space-time dependent linear responses.
Alternately, %we relate these topological terms
we relate these topological terms
to what are known as `anomalies' appearing in the theories describing the responses. 
Since the `anomalies' are known to be insensitive 
to whether the underlying fermions
are interacting or not, our so-obtained description of the
topological features demonstrates the insensitivity of 
these topological insulators 
(superconductors)
to 
adiabatic deformations by
interactions.

The general picture emerging from the results presented in this paper is that the topological insulators (superconductors) 
appearing in the ``ten-fold list'' can be viewed as generalizations of the $d=2$ Integer Quantum Hall Effect to systems 
in different dimensions $d$ and with different (``anti-unitary'') symmetry\cite{SRFL} 
properties. While the ``ten-fold classification scheme'' was originally established in \cite{SRFL,kitaevclassification}
for noninteracting fermions, the characterization in terms of anomalies implies that this extends also to 
all those {\it interacting} systems which can be adiabatically connected to noninteracting topological insulators 
(superconductors) without closing the bulk gap. 
(This may include fairly strong interactions, albeit typically not expected to exceed the noninteracting bulk gap.) 
One may expect that to any of the topological insulators (superconductors) in the ``ten-fold list'' 
(viewed as  generalizations of the Integer Quantum  Hall Effect)  corresponds a set of ``fractional'' 
topological insulators (superconductors) {\it  not} adiabatically connected to a noninteracting one, 
in analogy to the case of the two-dimensional Quantum Hall Effect. 
This includes, e.g., a recently proposed  three-dimensional ``fractional'' topological insulator
\cite{fractional}.
One expects a description in terms of anomalies to carry over to all such systems and to play a role in a 
(future) perhaps comprehensive characterization of such ``fractional'' topological insulators (superconductors). 
In the present paper, however, we focus on those interacting topological insulators (superconductors) which can be adiabatically connected to a noninteracting system of fermions.

Let us focus now on the topological insulators (superconductors) 
in $d=3$ spatial dimension (see also Table \ref{table1}).  
From a conceptual point of view it is the {\it surface responses}
which are simplest to describe and they are 
quantized 
(but they may not necessarily be most easily
accessible experimentally; therefore we also discuss the
{\it bulk responses} further below):   

\vskip .1cm

\noindent {\it Charge surface response}:
this is, in particular,  relevant for the 
(``$\mathbb{Z}_2$'', or ``spin-orbit'') topological insulator which is time reversal invariant. Upon
subjecting its surface to a weak time reversal symmetry breaking perturbation (in
the zero temperature limit),
the surface turns into a quantum Hall insulator whose electrical surface
Hall conductance 
takes on the 
quantized
%universal 
value
\cite{Comment-LFSG-PRB1994} 
\begin{equation}
\label{ElectricalHallConductance}
\sigma_{xy}/(e^2/h) = {n\over 2} 
\end{equation}
(a multiple of half the conductance quantum) as the strength of the symmetry breaking
perturbation is reduced to  zero (always at zero temperature).
Here
$n=0$ and $n=1$ for the ``$\mathbb{Z}_2$'' (or ``spin-orbit'') topological insulator
\cite{qilong,Comment-LFSG-PRB1994} 
(in the so-called symmetry class AII), in the topologically trivial
and non-trivial phase, respectively. 
While the surface of $\mathbb{Z}_2$ topological 
insulators in class AII may exhibit 
any odd (even) number Dirac cones in the topologically
non-trivial (trivial) phase at the microscopic level,
only the odd-even parity, $n=1$ and $n=0$ of that number
is topologically protected.
On the other hand, 
for the less familiar topological insulator
in symmetry class AIII,
it will become evident from the results
of the present paper that 
the number $n$ in the surface Hall conductance 
(\ref{ElectricalHallConductance})
is not restricted to $n=0$ or $1$
but 
the maximal (in magnitude) value it can take 
on is equal to
%can be as large as 
the topological charge (winding number) $\nu$
defined for the bulk topological states.
\cite{Pavan09,SRFLnewJphys}

\vskip .1cm
\noindent {\it Spin surface response}:
analogous effects are known\cite{diamondCI}
for the time reversal invariant
topological (spin-singlet) superconductor
in symmetry class CI in $d=3$ spatial dimension.  
Subjecting its surface, as above,  to a weak time reversal symmetry breaking 
perturbation (in the zero temperature limit),
the surface turns into what is known as the ``spin quantum Hall insulator''
\cite{REF-SQHE,footnoteNotConfuseQSH}.
Due to spin-singlet pairing this superconductor has $SU(2)$ Pauli-spin rotation symmetry
which permits the definition of the `surface spin conductivity'.
In particular\cite{REF-SQHE}, a gradient of magnetic field  within
the surface (say in the $z$-direction of
spin space) leads to a  spin-current perpendicular to the gradient (and within the surface).
This defines the `surface spin-Hall conductance' which takes on
%\cite{diamondCI}
the universal value
\begin{equation}
\label{UniversalSpinQuantumHallConductivity}
\sigma^{(spin)}_{xy}/ { (\hbar/2)^2\over h}
={n\over 2} 
\end{equation}
($n$-times half the `spin-conductance quantum' 
${ (\hbar/2)^2\over h}$, 
where 
the maximal (in magnitude) value 
the integer $n$ can take 
on is equal to
%$n$ can be as large as 
the integer $\nu$ from the $\mathbb{Z}$-classification
\cite{SRFL,diamondCI}
of the topological superconductor in symmetry class CI)
as the time reversal symmetry breaking perturbation is reduced to zero.

\vskip .1cm
\noindent {\it Thermal surface response}:
as we show in section \ref{subsec:gravitational instanton term}
of this paper, an analogous effect occurs for the thermal response 
at the surface of the time reversal invariant topological superconductor
in symmetry class DIII in $d=3$ spatial dimensions:
subjecting its surface, as above, to a weak time reversal symmetry breaking perturbation 
(in the low temperature limit),
a temperature gradient within the surface leads to a  heat (energy)
current in the perpendicular direction in the surface. The so-defined surface thermal Hall 
conductance 
$\sigma^T_{xy}$
(when divided by temperature) tends, as
will become evident from the results of the present paper, 
in the zero-temperature limit to a quantized value
\begin{equation}
\label{UniversalThermalHallConductivity}
(\sigma_{xy}^T/T)/ {(\pi k_B)^2 \over 3h}
= \pm c/2, \ \ {\rm where} \ \ c=n/2
%(\sigma^T_{xy}/T)/{\pi\over 6} (k_B^2/h)^2
%= {n \over 2}  
\end{equation}
where 
the maximal 
(in magnitude) value the integer $n$ can take on is equal to
%$n$ can be as large as is 
the integer $\nu$
from the $\mathbb{Z}$-classification 
\cite{SRFL,SRFLnewJphys} 
of the topological
superconductor in symmetry class DIII,
as the symmetry breaking perturbation is reduced to zero.
[$c \times {(\pi k_B)^2 \over 3h}$ 
is the thermal conductance for a Majorana fermion when $c=1/2$ 
(its central charge).]

These surface Hall conductances provide a characterization
of the mentioned topological insulators, irrespective of whether electron interactions
are present or not:
if we start out with a non-interacting topological insulator, one
can explicitly compute the
theory describing various space-time dependent responses.
(For the thermal responses of the DIII topological superconductor
in $d=3$ spatial dimension, this is done in
Subsec.\ \ref{subsec:gravitational instanton term}
of this paper. 
For the $SU(2)$ spin-responses of the topological singlet superconductor
in symmetry class CI this was done in Ref. \onlinecite{diamondCI}.
For a significant part of the list of all
topological insulators (superconductors) this
is done more generally in Section \ref{ChapterAnomalies} of this paper
for all dimensionalities.)
%Because 
Owing to the fact that
the underlying insulators are topological,
the field theories for the responses
turn out to be described by what are called anomalies.
These anomalies describe the responses both,
in the bulk and at the surface. 
The charge, spin, and thermal surface responses
are examples.
\cite{footnote-ChernSimons-Anomaly}
Anomalies are known
to be insensitive to the presence or absence of
interactions.
For this reason, the 
(maximal value of the)
dimensionless charge, spin, and thermal surface responses
are independent of the strength of the
interactions. These surface responses can only change
when a bulk quantum phase transition is crossed
(at which the bulk gap closes).

While these surface responses
are quantized
and theoretically useful in that
they permit one to understand
the stability of the topological insulator (superconductor)
phases to interactions (for the cases discussed above,
and in Section
\ref{ChapterAnomalies} for general dimensionalities),
they may not 
all be
directly accessible experimentally.
Therefore we discuss below also the various {\it bulk} responses.

The {\it bulk} responses that we
find are of three types: {\it charge} response, previously shown to lead to  a quantized ${\bf E} \cdot {\bf B}$ term in the ordinary $\mathbb{Z}_2$ 
topological insulator (``axion electrodynamics'')~
\cite{wilczekaxion,qilong,essinmoorevanderbilt}; 
{\it gravitational} response, when energy flows lead to an analogue of this term for gravitational fields, leading to a Lense-Thirring frame-dragging effect 
\cite{LenseThirring}
when a temperature gradient is applied; 
and {\it magnetic dipole} response, 
when a magnetic dipole current induced by an applied perturbation 
leads to an electrical field.  
A single phase may show more than one of these effects; for example, a phase with a conserved $SU(2)$ spin current can show a non-Abelian 
response of this type
in the presence of an $SU(2)$ gauge field coupling to this current, but 
will also show a magnetic dipole response
via its coupling to ordinary $U(1)$ electromagnetism.  
We obtain these possible responses
for each of the five symmetry classes in three dimensions supporting topological phases.
\cite{SRFL,kitaevclassification}
As in the classification in Ref.~\onlinecite{SRFL}, 
the approach we take is based upon the 
surfaces of these topological phases; these surfaces carry currents leading to new terms in the effective action of gravitational and electromagnetic fields.
Our results for the various symmetry classes with topological invariants in three dimensions are summarized in Table \ref{table1}.

\begin{table}
\begin{tabular*}{3 in}{@{\extracolsep{\fill}}  c  c  c c }
Symmetry&Charge&Gravitational
&Dipole\cr
\hline\hline\\
AII&\checkmark&\checkmark& \\
CI&&\checkmark&\checkmark \\
CII&&\checkmark&\checkmark \\
DIII&&\checkmark&\\
AIII&$^*$&\checkmark&$^*$
%2 \log (2) / 3 \approx

\end{tabular*}
\caption{
\label{table1}
Electromagnetic and gravitational (thermal) responses for
five out of ten Altland-Zirnbauer symmetry classes 
(AII, CI, CII, DIII and AIII).  
The assumptions made in the first four classes are 
that $U(1)$ conserved currents arise from electrical charge 
and that $SU(2)$ conserved currents arise from spin.  
$^*$: 
In class AIII, 
the $U(1)$ conservation law may arise either 
from charge or one component of spin.}
\end{table}

These bulk  responses are
 ``topological'' to varying degrees.  
The charge response is topological both in its spatial dependence and as a term of the effective action: quantization of the response is tied to quantization of the electrical charge and the Dirac quantization condition.  
The gravitational response is topological in terms of the spatial dependence, but its coefficient is related to the mass or energy of the underlying particles and hence not quantized to the same degree as the charge response.
The magnetic dipole response
is not topological in the sense of being metric-independent, but it does arise from sample boundaries in the same way as the other responses.

This paper is organized as follows: 
We begin in Sec.\ \ref{Charge responses}
by reviewing the axion electromagnetism 
for the three-dimensional topological insulators in 
the spin-orbit symmetry class (symmetry class AII). 
In Sec.\ \ref{Gravitational responses},
thermal response of three-dimensional time-reversal
invariant topological superconductors (such as 
the B-phase of $^{3}\mathrm{He}$)
is discussed by exploiting a close analogy of 
electromagnetism and gravity in Newtonian approximation. 
In Sec.\ \ref{Dipole responses}, 
dipole response is discussed for three-dimensional topological
phases when at least one component of spin is conserved. 
All these responses will be discussed from much a 
broader
perspective
in Sec.\ \ref{ChapterAnomalies} in terms of anomalies of various kind
(chiral anomaly, gauge anomaly, gravitational anomaly),
and the descent relation relating these anomalies.
We conclude in
Sec.\ \ref{Conclusions}.

\section{Charge responses}
\label{Charge responses}

\begin{figure}[tb]
\begin{center}
    \includegraphics[width=.3\textwidth]{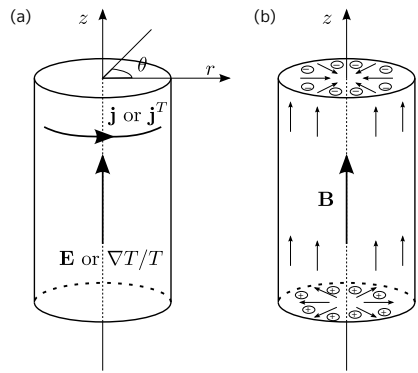}
  \caption{
Electric and thermal response of 
topological insulators, 
and thermal response 
of topological triplet superconductors, 
in a cylindrical geometry. 
(a): 
Electric ($\mathbf{j}$) or thermal ($\mathbf{j}^T$) 
current driven by
applied electric field ($\mathbf{E}$) or thermal gradient 
($\boldsymbol{\nabla} T/T$). 
(b):
A response dual to (a)
where an applied magnetic field in $z$-direction
induces charge polarization. 
\label{cylinder fig}
}
\end{center}
\end{figure}

For an explicit example, consider a cylinder of topological insulator with surface Hall conductance 
$\pm e^2/(2 h)$, defined with reference to the outward normal
(see Fig.\ \ref{cylinder fig}).
(Below, we choose plus sign for the surface Hall conductance, 
by subjecting the surface to a weak external time-reversal symmetry source.)
The motivation for considering this example in some detail is that it will lead to a direct interpretation of the corresponding gravitational response below.  The current response to an applied electrical field along the cylinder axis is
\beq
{\bf j} = j_{\theta} \ {\bf  \hat \theta}
\quad 
{\rm where} 
\quad 
j_\theta = {e^2 \over 2 h} E_z.
\eeq
Now the magnetic field induced by this current follows from one of Maxwell's equations,
\beq
\boldsymbol{\nabla} \times {\bf B} = {4 \pi \over c} {\bf j},
\eeq
which leads to the magnetostatic equation
\beq
{\bf B}({\bf x}) = {1 \over c} \int {\bf j}({\bf x}^\prime) \times {({\bf x} - {\bf x}^\prime) \over |{\bf x} - {\bf x}^\prime|^3}\,d^3 x^\prime.
\eeq
The result for a thin cylinder is that the magnetic field at the cylinder axis, 
well away from the cylinder ends, is given by
${\bf B} = B_z {\bf \hat z}$ with 
\beq
B_z  = {1 \over c} \int_{-\infty}^{\infty} {r (2 \pi r) j_\theta \over (r^2 + a^2)^{3/2}}\,da = {4 \pi \over c} j_\theta = {2 \pi e^2 E_z \over h c}.
\eeq
This magnitude follows from minimizing the magnetic energy
\beq
H_B = {B^2 \over 8 \pi} - {e^2 \over 2 h c} {\bf E} \cdot {\bf B}
\eeq
%for the coupling $\theta = -\pi$ in 
which follows from the Maxwell Lagrangian supplemented with the theta term
(axion term)
\begin{eqnarray}
\mathcal{L}_{\theta}
=
\frac{\theta e^2}{2\pi h c} {\bf E}\cdot {\bf B} 
=
\frac{\theta e^2}{16\pi h c} 
\epsilon^{\mu\nu\rho\lambda}
F_{\mu\nu}F_{\rho\lambda}
\end{eqnarray}
for the coupling $\theta = -\pi$.
%(\lookhere).  
(The negative sign in this equation is picked out by the choice of the direction of the current flow around the cylinder.)

To understand the dual response, which is an electrical field induced by an applied magnetic field, one needs to include the ends of the cylinder.  Applying a magnetic field normal to a Hall layer increases or decreases the charge density depending on the direction of the field, as is required for the charge continuity equation to follow from Maxwell's equation
\beq
{\partial {\bf B} \over \partial t} + 
\boldsymbol{\nabla} \times {\bf E} = 0.
\eeq
Hence an applied magnetic field induces an electrical polarization along the interior of the cylinder.  We now turn to a gravitational version of the above physics, generated by energy flows from surface thermal Hall layers.

\section{Gravitational responses}
\label{Gravitational responses}

%%%%%% BEGIN FIGURE
%%%%%% END FIGURE

\subsection{gravitoelectromagnetism} 

Our approach will be to start from the energy flow at surfaces of a topological phase, which is the microscopic source of the gravitational response.  The importance of this response is that it is the only one that exists in the important symmetry class DIII, which includes superfluid $^3$He.  We use this phase as an explicit example in the following.  The surface Majorana mode that exists in this phase does not carry charge, but does carry heat, leading to a thermal Hall effect.  Hence a temperature gradient applied to a cylinder leads to an energy flow perpendicular to the applied gradient 
%(Fig.~\lookhere):
\beq
j^T_\theta = \sigma_{xy}^T (- \partial_z T) = c^{-2} T \sigma_{xy}^T 
E_{g,z},
\eeq
where for future use we have treated temperature as a scalar potential generating a field ${\bf E}_g = -c^2 (
\boldsymbol{\nabla} T) / T$ 
with units of acceleration.  
The physical meaning of this scalar potential was worked out 
by Luttinger in his derivation of the thermal transport coefficients
\cite{Luttinger1964}:
in a near-equilibrium system, the effect of a thermal gradient is equivalent to that obtained from a gravitational potential $\psi$ such that
\beq
\boldsymbol{\nabla} \psi = {{\boldsymbol{\nabla} T} \over T},
\eeq
where $\psi$ is the gravitational potential energy per mass, divided by $c^2$.

This rotational energy flow couples to the gravitational field at the first post-Newtonian approximation (i.e., the coupling is down by a factor $v/c$ compared to the static gravitational effect present in the absence of the applied gradient).  Because temperature couples to the local energy density in the same way as an applied gravitational potential, as used by Luttinger in his derivation of the thermal Kubo formula
\cite{Luttinger1964}, we can view this effect similarly to the charge response above, as a gravitational ``magnetic'' field resulting from the energy flow that was induced by a gravitational ``electric'' field reflecting the temperature gradient.

This analogy can be made precise in the near-Newtonian limit using the gravitoelectromagnetic equations
\cite{ClarkTucker2000} that apply to a near-Minkowski metric.  The relevant equation is that a mass current induces a gravitomagnetic field $B_g$, defined more precisely below, via the equation
\beq
\boldsymbol{\nabla} \times {\bf B}_g = {-4 \pi G {\bf j}_m \over c}.
\eeq
Here ${\bf j}_m$ is the (three-dimensional) mass current density, 
satisfying ${\bf j}_m = {\bf j}^T / c^2$, and $G$ is 
the effective Newton constant of the material. 
The negative sign in this equation compared to the corresponding Maxwell's equation is physically significant and results from the difference that equal masses attract, while equal charges repel.  The field $E_g$, like $B_g$, has units of acceleration, and the gravitational force on a test particle of 
small mass $m_{test}$ is
\beq
{\bf F} = m_{test} 
\left({\bf E}_g + 2 {{\bf v} \over c} \times {\bf B}_g
\right),
\eeq
where ${\bf v}$ is the particle velocity.  The factor of 2 here results from the spin-2 nature of the gravitational field.

Now, by the same steps as above, there is an induced field along the cylinder axis
\beq
B_g = {4 \pi G {\bf j}^T_\theta \over c^3} = {4 \pi G \over c^3} {T \sigma_{xy}^T {E_g}_z \over c^2}.
\eeq
Since $\sigma_{xy}^T$ has the units  $k_B^2 T / h$ of a two-dimensional thermal conductivity, the ratio between $B_g$ and $E_g$ is of the form $G ({\rm energy}^2) / (h c^5)$,  which is dimensionless (the gravitational analogue of the fine structure constant that appears in the charge case).

The gravitomagnetic field then has exactly the same spatial dependence as the magnetic field in the axion case computed above.  In particular, it is topological (e.g., the field at the cylinder axis does not fall off as the cylinder radius becomes larger) and scales with the energy flow, which in turn scales quadratically with the mass of the underlying particles.

\subsection{gravitational instanton term}
\label{subsec:gravitational instanton term}

We now discuss the gravitational response 
in topological insulators and superconductors
from more formal point of view. 
When discussing electromagnetic responses in 
topological insulators, 
we can couple electrons
to an external 
(background)
$U(1)$ gauge field.
The $\theta$-term 
in the effective action for the gauge field
then results by integrating over the gapped electrons.
In order to discuss gravitational and thermal responses,
we can take a similar approach:
we can introduce an external gravitational field 
that couples to fermions 
(electrons for topological insulators,
and fermionic Bogoliubov quasiparticles for
topological superconductors.)
By integrating  over the gapped fermions,
we obtain
an effective gravitational action.
The derivation of the effective action 
proceeds in a way quite  parallel to that of the
$U(1)$ case: Indeed, both of them are related to 
a chiral anomaly as we will see below. 

For topological insulators or superconductors
defined on a lattice, 
it is not obvious how
to couple fermions to gravity in a way fully
invariant under general coordinate transformations.
Also, there is of course
no Lorentz symmetry on a lattice. 
Yet, energy and momentum are conserved,
and one can think of introducing an external field which
couples to these conserved quantities. 
%The gravitoelectromagnetic approach 
%discussed in the previous subsection
%is not generally covariant 
The gravitoelectromagnetic approach 
discussed in the previous subsection
is based on a particular background (flat Minkowski metric), 
and 
is an approximation of the full Einstein gravity 
in the limit where the mass flows are small in some particular
reference frame defined by the system with no thermal perturbation.  
%and assumes that the mass flows are small in some particular
%reference frame defined by the system with no thermal perturbation.  

However, all topological insulators (superconductors)
are known\cite{SRFLnewJphys} to possess a representative
in the same topological phase which is described
by a Dirac hamiltonian.
Fermions whose dynamics is described by a Dirac
hamiltonian can naturally be coupled to a gravitational
background field. (The theory is fully
Lorentz invariant, and the coupling to gravity
is fully invariant under general coordinate transformations,
and can be described in terms of the spin connection.)
For this reason we provide (below) a derivation of the effective
action in terms of the Dirac representative of the topological
phases.
The
topological features of the effective
action for the gravitational responses are expected to
be independent of the choice of representative
in the topological class,
and thus
to have a much more general applicability.
Physically, such gravitational responses describe thermal response functions.
\cite{Luttinger1964}

We thus consider the following single $4\times4$ continuum Dirac model, 
\begin{eqnarray}
H
=
\int d^3 x\,
\psi^{\dag}
\left(
-{i} \boldsymbol{\partial} \cdot\boldsymbol{\alpha}+m\beta
\right)
\psi,
\end{eqnarray}
where $\psi^{\dag}$ and $\psi$ represent creation/annihilation
operator of complex fermions, and 
$
\boldsymbol{\alpha}
=
\sigma_1 \otimes \boldsymbol{\sigma}
$
and
$
\beta = \sigma_3 \otimes \sigma_0
$
are the Dirac matrices
($\sigma_{0,1,2,3}$ are 
standard Pauli matrices).
(In this subsection, we use 
natural units, 
$c=\hbar=1$,
and set the Fermi velocity to be one for simplicity.) 
For topological superconductors, 
we need to use real (Majorana) fermions instead 
of complex fermions. 

We assume the Dirac model is in a topologically non-trivial phase 
for $m>0$ while it is in a trivial phase for $m<0$:
While this does not look apparent from the action in
the continuum limit, 
when the Dirac model is derived from an appropriate lattice model,
the sign of the mass does determine the nature of the phase. 
In the presence of a gravitational background,
the fermionic action is given by
\cite{nakahara} 
\begin{eqnarray}
&&
S [m,\bar{\psi},\psi,e]
=
\int  d^4 x\,
\sqrt{g} 
\mathcal{L},
\\%%%%%
&&
\mathcal{L}
=
\bar{\psi}
e^{\ }_{a}{ }^{\mu}
{i}
\gamma^{a}
\Big(
\partial_{\mu}
-
\frac{{i}}{2}
\omega_{\mu}{ }^{ab}
\Sigma_{ab}
\Big)
\psi 
-
m
\bar{\psi}
\psi,
\nonumber
\end{eqnarray}
where 
$\mu,\nu,\ldots=0,1,2,3$
is the space-time index, 
and
$a,b,\ldots=0,1,2,3$
is the flat index;
$e_{a}{ }^{\mu}$ is 
the vielbein,
and $\omega_{\mu}{ }^{ab}$ is 
the spin connection;
$\Sigma_{ab}=\left[\gamma_a,\gamma_b\right]/(4 i)$. 
(See Ref.\ \onlinecite{footnote_reimann} for our conventions
of metric, vielbein, spin connection, etc.)
The effective gravitational action $W_{\mathrm{eff}}[m,e]$
for the gravitational field is then obtained from 
the fermionic path integral
\begin{eqnarray}
e^{{i} W_{\mathrm{eff}}[m,e]}
=
\int\mathcal{D}\left[\bar{\psi},\psi\right]
e^{{i} S[m, \bar{\psi},\psi,e]}
\end{eqnarray}

A key observation is that the continuum 
Hamiltonian $H$
enjoys a continuous chiral symmetry:
we can flip the sign of mass, in a continuous fashion,
by the following chiral rotation 
\begin{eqnarray}
\psi\to\psi=e^{{i}\phi \gamma_{5}/2}\psi',
\quad
\psi^{\dag} \to \psi^{\dag}=
\psi^{\dag \prime} e^{-{i}\phi\gamma_{5}/2},
\end{eqnarray}
under which 
\begin{eqnarray}
\bar{\psi}\left(i \partial_{\mu}\gamma_{\mu}-m\right)\psi
=
\bar{\psi}'\left(i \partial_{\mu}\gamma_{\mu}-m'(\phi)\right)\psi',
\nonumber \\%%%%%
m'(\phi)
= 
me^{{i}\alpha\gamma_{5}}=m\left[\cos\phi+{i}\gamma_{5}\sin\phi \right],
\end{eqnarray}
so that $m'(\phi=0)=m$ and $m'(\phi=\pi)=-m$.
Since $m$ can continuously be
rotated into $-m$, one would think, naively, 
$W_{\mathrm{eff}}[m,e]=W_{\mathrm{eff}}[-m,e]$.
This naive expectation is, however, not true because of chiral anomaly.
The chiral transformation which rotates $m$ continuously
costs the Jacobian $\mathcal{J}$ of the path integral measure, 
\begin{eqnarray}
\mathcal{D}\left[\bar{\psi},\psi\right]
=
\mathcal{J}\mathcal{D}\left[\bar{\psi'},\psi'\right]. 
\end{eqnarray}
The chiral anomaly (the chiral Jacobian $\mathcal{J}$) is responsible
for the $\theta$-term.
%\begin{eqnarray}
%\mathcal{J}
%\!\!&=&\!\!
%\exp\left(
%- 2 {i} \int d^d x\, \alpha(x)
%\sum_n \varphi^{\dag}_n(x) \gamma_5 \varphi^{\ }_n(x)
%\right)
%\end{eqnarray}
The Jacobian $\mathcal{J}$ can be computed explicitly by 
the Fujikawa method\cite{fujikawa}
with the result 
%(for symmetry class DIII)
% as 
\begin{eqnarray}
\label{GravitationalInstantonTerm}
W^{\theta}_{\mathrm{eff}}
\!\!&:=&\!\!
-\ln \mathcal{J}
\\%%%%%
\!\!&=&\!\!
%\frac{  \theta}{2\times 384\pi^2} 
\theta \ 
\frac{1}{2} \ 
\left [
\frac{ 1}{2\times 384\pi^2} 
\int d^4 x
\sqrt{g} \epsilon^{cdef}
R^{a}{ }_{b cd }
R^{b}{ }_{a ef}
\right ],
\nonumber
\end{eqnarray}
when $m>0$ while $W^{\theta}_{\mathrm{eff}}=0$ when $m<0$.
The expression in square brackets is the so-called Dirac genus (see
Section \ref{ChapterAnomalies} below for details)
which is equal \cite{nakahara}, by the Atiyah-Singer index theorem,
to the index of the Dirac operator in the curved background.
The multiplicative prefactor 1/2
arises because of the
Majorana nature of the Bogoliubov quasiparticles.
%and is absent if instead we consider complex (Dirac) fermions. 
%We note that for the case of the Bogoliubov quasiparticles in
%topological superconductors, in particular 
%for the topological superconductors (superfluids) 
%in symmetry class DIII considered here (including ${}^{3}$He B),
%the whole
%effective action in Eq.  \ref{GravitationalInstantonTerm}
%has acquired a multiplicative prefactor of 1/2 because of the
%Majorana nature of the Bogoliubov quasiparticles.
%This 
The index in square brackets 
is in fact an even integer (by Rochlin's Theorem\cite{RochlinTheorem}).
Therefore, (1/2) of  that expression, i.e. half the index,
is an integer.  Thus the gravitational effective action
$W^{\theta}_{\mathrm{eff}}$ 
in Eq.\ (\ref{GravitationalInstantonTerm})
equals 
$\theta$ times
an integer, i.e. it is
a so-called $\theta$-term.
Now, since $\theta \to -\theta$ under
time reversal, the theta angle is fixed by
time reversal symmetry and periodicity to either $\theta=0$
or $\theta=\pi$. The former corresponds
to a topologically trivial state, and
$\theta=\pi$ to the topologically non-trivial
state.
(See, for a similar discussion on
the derivation of the 
$\theta$-term,
i.e., ${\bf E}\cdot {\bf B}$ term, 
for the electromagnetic response, 
Ref.\ \onlinecite{Pavan09},
as well as for the non-Abelian $SU(2)$ response
in Ref.\ \onlinecite{diamondCI}.)
Note that if instead we consider complex (Dirac) fermions
in the background gravity field, 
the theta angle $\theta$ is an integer multiple of $2\pi$,
but not of $\pi$ as in the Majorana case.

The part of the effective action, 
which is not related to the Fujikawa Jacobian, 
takes the form of the Einstein-Hilbert action
$W_{\mathrm{EH}}  =
%(16\pi G_N)^{-1}
(16\pi G)^{-1}
\int d^4 x\,
\sqrt{g} 
R 
$
where $G$ is the effective Newton constant
in the bulk of the topological insulator (superconductor). 
The gravitoelectromagnetism equations mentioned above can be 
derived from the effective action by taking the Newtonian limit
(near Minkowski limit). 
%The same calculation can be repeated for
%We note that for the case of the Bogoliubov quasiparticles in
%topological superconductors, in particular 
%for the topological superconductors (superfluids) 
%in symmetry class DIII considered here (including ${}^{3}$He B),
%the whole
%effective action in Eq.  \ref{GravitationalInstantonTerm}
%has acquired a multiplicative prefactor of 1/2 because of the
%Majorana nature of the Bogoliubov quasiparticles.

To make the connection with the existence of
topologically protected surface modes we note that
when there are boundaries (say) in the $x^3$-direction
at $x^3=L_+$ and at $x^3=L_-$, 
the gravitational instanton term
$W^{\theta}_{\mathrm{eff}}$,
at the non-trivial time-reversal invariant
value $\theta=\pi$ of the angle $\theta$,
can be written 
in terms of the gravitational Chern-Simons terms at the boundaries,  
\begin{eqnarray}
W^{\theta}_{\mathrm{eff}}
= 
I_{\mathrm{CS}}|_{x^3=L_+}
-I_{\mathrm{CS}}|_{x^3=L_-}, 
\end{eqnarray}
where ($i,j,k=0,1,2$)
\begin{eqnarray}
I_{\mathrm{CS}}
=
\frac{1}{2}  
\  
\frac{1}{4\pi}
\ 
\frac{c}{24}
\int d^3 x
\epsilon^{ijk} 
\mathrm{tr}
\Big(
\omega_{i} \partial_{j} \omega_{k}
+
\frac{2}{3}
\omega_{i}
\omega_{j}
\omega_{k}
\Big).
\end{eqnarray}
with $c=1/2$.
This kind of relationship between
the theta term 
and 
the Chern-Simons type term in one lower dimension
is a special case of the so-called descent relation
and will be discussed further in
Section \ref{ChapterAnomalies}.
This value of the coefficient of the gravitational Chern-Simons term
is one-half of the canonical value 
$
(1/4\pi)
\times
(c/24)
$
with $c=1/2$.
As before, for fermions with a reality condition (Majorana fermions),
the canonical value of the coefficient of the gravitational Chern-Simons term 
corresponds to $c=1/2$, as opposed to $c=1$ for fermions without
a reality condition 
(and $c=n/2$ for $n$ species of fermions with a reality condition.) 
As discussed by
Volovik\cite{Volovik90}
and, 
Read and Green\cite{readgreen}
in the context of the two-dimensional chiral
$p$-wave superconductor, the coefficient of the gravitational Chern-Simons term
is directly related to the thermal Hall conductivity, which in our case 
is carried by the topologically protected surface modes\cite{FootnoteCentralChargeEdge}.
(See Eq. (\ref{UniversalThermalHallConductivity}) of the Introduction.)

\section{Dipole responses}
\label{Dipole responses}

\subsection{topological singlet superconductor (class CI)
and spin chiral topological insulator (class CII)} 

The last response we consider can be measured in systems 
with a conserved spin or magnetic dipole current.  
Among the five 
symmetry classes which admit 
a topological phase in three-spatial dimensions, 
we thus focus on 
topological singlet superconductors in 
symmetry class CI 
(possessing time-reversal and spin rotation invariance),
and also on 
topological insulators in 
symmetry class CII 
(possessing time-reversal but without spin rotation invariance)
(see Table \ref{table1}).

Simple lattice models of the three-dimensional topological singlet superconductor
in symmetry class CI
were discussed previously 
on the diamond lattice  
\cite{diamondCI}
and on the cubic lattice
\cite{Pavan09},
for which, in the presence of a boundary (surface), 
there is a stable and non-localizing Andreev bound state. 
Similar to the quantized ${\bf E}\cdot {\bf B}$ term for
the charge response in the topological insulator,
the response 
of topological singlet superconductors
to a fictitious external $SU(2)$ gauge field
(``spin'' gauge field which couples to conserved spin current)
is described by
the $\theta$-term at $\theta=\pi$ in
the (3+1) dimensional $SU(2)$' Yang-Mills theory.
\cite{diamondCI}
The $\theta$-term predicts the surface quantum Hall effect
for spin transport (the spin quantum Hall effect),
as already mentioned in the Introduction 
(Section \ref{Label-Introduction}).

To detect such a quantum Hall effect for the $SU(2)$ symmetric spin current 
requires 
a fictitious external spin gauge field, 
and hence one would think it cannot be detected experimentally.
Nevertheless, we discuss in this section that the 
electromagnetic
response carried 
by the dipole moment
of the spin current can be 
measurable.
(See Ref.\ \onlinecite{Goryo2008} for a similar discussion
on the dipole response in a $^{3}$He-A superfluid thin film
or two-dimensional $p$-wave paired states. )

The topological insulator in symmetry class CII
(called ``spin chiral topological insulator'' in Ref.\ \onlinecite{Pavan09}) 
is in many ways analogous to the more familiar
quantum spin Hall effect in two spatial dimensions, 
but requires
the chiral symmetry in addition to time-reversal symmetry.
(For a lattice model of the $\mathbb{Z}_2$ 
topological insulator in symmetry class CII, 
See Ref.\ \onlinecite{Pavan09}). 
Just like an intuitive understanding of 
the quantum spin Hall effect can be obtained by starting from 
two decoupled and independent quantum Hall systems  
with opposite chirality for each spin and then glue them together,
this spin chiral topological insulator can be obtained by
considering two independent topological insulators in symmetry class AIII. 
More general quantum spin Hall states or spin chiral topological insulators
can then be obtained by destroying the $S_z$ conservation 
by mixing spin up and down components. 
The dipole response for class CII topological insulators,
which we will describe below,
assumes that a $U(1)$ part of the $SU(2)$ spin rotation symmetry is 
conserved (i.e., one-component of spin is conserved).
However, even when there is no such symmetry, if 
mixing between two species is weak, we can still have such 
a dipole response.

\subsection{magnetic dipole responses}

The spin current response at the surface of such a system 
to an applied magnetic field ${\bf B}$ via the Zeeman effect can be written as
\beq
j^a_i = \alpha \epsilon_{ijk} (\partial_j \theta)\partial_k B_a,
\eeq
where $\alpha$ is some constant. 
Here we have introduced a scalar field $\theta$ (``axion'' field)\cite{diamondCI}, 
by analogy with the local electromagnetic polarizability of 
the (AII, spin-orbit)
topological insulator, 
to describe the spatial location of the dipole current, 
which as before is a surface property.  
Here $j^a_i$ represents the $a$-th component of a magnetic dipole current of dipoles 
in spatial direction $i$.  
Such a current can generate two types of {\it static} electromagnetic responses: a dipole density through the continuity equation
\beq
\partial_i j^a_i + \partial_t n^a = 0,
\eeq
and an electrical field through the equation
\beq
\left(
\boldsymbol{\nabla} \times {\bf E}\right)_i 
= \epsilon_{ijk} \partial_j E_k = {\mu \over 4 \pi} \partial_a j^a_i, 
\label{formuladipole}
\eeq
where $\mu$ is the permeability of the material of interest.
(One could alternately have a time-varying magnetic field, just as a current density can produce either a constant magnetic field or a time-varying electrical field.)  The second response may be unfamiliar but can be derived from elementary principles; 
see Ref.~\onlinecite{MeierLoss2003} for a discussion of how it can be measured experimentally.  Start from a dipole field in the lab frame; take one copy with the dipoles pointing along some direction ${\bf \hat n}$ and boost that along ${\bf v}$, and take another copy with the dipoles pointing along $-{\bf \hat n}$ and boost that along $-{\bf v}$.  For a dipole density $n^a$, 
this leads, in the comoving frame,  to the field
$B_a = (\mu/4 \pi) n^a$, 
and hence
\beq
\boldsymbol{\nabla} \cdot {\bf B} = {\mu \over 4 \pi} \partial_a n^a.
\eeq
%where $\mu$ is the permeability of the material of interest. 
Using the non-relativistic Lorentz transformation law
\beq
{\bf E}
\to 
\gamma \left({\bf E} +  {\bf v} \times {\bf B}\right)
%{\bf E}^\prime_\perp = \gamma \left({\bf E} +  {\bf v} \times {\bf B}\right)_\perp
\eeq
with $\gamma\simeq 1$
leads to Eq.\ (\ref{formuladipole}), with $j^a_i = v_i n^a$.
%:
%\beq
%E_i = \epsilon_{ijk} v_j B_k, \quad (\nabla \times {\bf E})_i = \ldots.
%\eeq

Now we consider these responses for 
the surface spin current of 
a three-dimensional topological singlet superconductor. 
The spin Hall current is always divergence-free by commutation of derivatives,
\beq
\partial_i j^a_i = \alpha \epsilon_{ijk} \partial_i 
\left(\partial_j \theta \partial_k B_a \right) = 0,
\eeq
since whichever term the $\partial_i$ acts on gives zero.
However, the electromagnetic response can be nonzero:
\beq
\epsilon_{ijk} \partial_j E_k = 
{\mu \over 4 \pi} \partial_a j^a_i = 
{\mu \alpha \over 4 \pi} 
\partial_a 
\left( \epsilon_{lmn} \partial_m \theta \partial_n B_a \right).
\eeq
There are two parts to this: one ``monopole'' part is only nonzero if $\partial_a B^a \not = 0,$ and we therefore neglect it.  There is also a term
\beq
{\mu \alpha \over 4 \pi} \epsilon_{lmn} (\partial_a \partial_m \theta) 
\partial_n B_a.
\eeq

%%%%%% BEGIN FIGURE
\begin{figure}[tb]
\begin{center}
    \includegraphics[width=.4\textwidth]{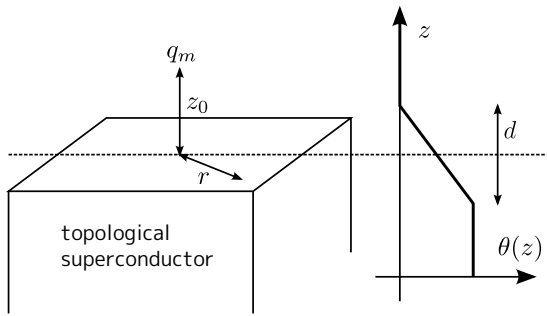}
  \caption{
Surface of a spin chiral topological insulator (class CII)
or topological singlet superconductor (class CI). 
\label{cylinder}
}
\end{center}
\end{figure}
%%%%%% END FIGURE

\subsection{example} 

As an example, 
we compute this response for the case of 
a surface of a topological singlet superconductor,
where the theta angle $\theta$ varies 
as a function of the distance from the surface
(Fig.\ \ref{cylinder}).
 For the response to be nonzero, we need $a=m=z$, so the response is to the $z$ component of the magnetic field.  We get, up to a possible sign,
\beq
\left( 
\boldsymbol{\nabla} \times {\bf E}\right)_x = - {\alpha \mu \over 4 \pi} \partial_z^2 \theta \partial_y B_z, \quad
\left( 
\boldsymbol{\nabla} \times {\bf E}\right)_y = {\alpha \mu \over 4 \pi} \partial_z^2 \theta \partial_x B_z.
\eeq
For the case where  $\theta$ is first  constant, then
changes linearly in $z$ within a  surface surface  layer,
and is then constant again outside this layer
(Fig.\ \ref{cylinder}), 
this response will occur entirely 
at the top and bottom surfaces of the region of linear change.  
As an example relevant to possible experiments, we compute this response for the magnetic field produced by a magnetic monopole field of strength $q_m$ (i.e., from one end of a long magnetic dipole), suspended a distance $z_0$ above a spin Hall surface layer where $\theta$ changes linearly across a thickness $d$.  
This surface layer gives two surfaces with
\beq
(\boldsymbol{\nabla} \times {\bf E})_x 
= j^m_x = \mp \beta \partial_y B_z,
\quad  
(\boldsymbol{\nabla} \times {\bf E})_y = j^m_y = 
\pm \beta \partial_x B_z.
\eeq
where  $\beta = (\alpha \mu)/(4 \pi) \ \pi/d$.
At the top layer,
the $z$ component of magnetic field is, in cylindrical coordinates,
\beq
B_z = {q_m z_0 \over (r^2 + z_0^2)^{3/2}},
\eeq
which leads to a surface magnetic current of magnitude
\beq
j^m_\theta = {3 \beta q_m z_0 r \over (r^2+z_0^2)^{5/2}}. 
\eeq
at the top surface.
Since
\beq
{\bf E}({\bf r}) = \int\,d^3{\bf r}^\prime {({\bf r}-{\bf r}^\prime) \times {\bf j}({\bf r^\prime)} \over |{\bf r}-{\bf r}^\prime|^2},
\eeq
we obtain that the electrical field from the top surface, at a height $z_1$ above the top surface (and directly above or below the original monopole), is
\beq
E_z(z_1) 
= \int_0^\infty\,(2 \pi r)\,dr\,{3 \beta q_m z_0 r \over (r^2+{z_0}^2)^{5/2}} {r \over r^2+{z_1}^2}.
\eeq
Evaluating this at the original height $z_0$ gives
\beq
E_z(z_0) = (6 \pi \beta q_m z_0) {2 \over 15 {z_0}^4} = {4 \pi \beta q_m \over 5 {z_0}^3}.
\eeq
Comparing this to the case of an image charge above a metal, 
we see that the electrical field falls off by one more power of height.  From the above, the dipole currents are localized to the top and bottom surfaces of the region where $\theta$ changes.  The bottom surface contributes with an opposite sign and with $z \rightarrow z+d$, so we obtain
\beq
E_z(z_0) = {4 \pi \beta q_m \over 5} \left({{z_0}^{-3} - (z_0+d)^{-3}}\right),
\eeq
so that for $d \ll z_0$ the electric field falls off as the fourth power of distance.

We can understand the scaling of the result by noting that $q_m$ divided by length cubed has units of magnetic field per length; multiplying by $\beta$ converts this to a 2D magnetic charge current density, which has the same units as electric field.  While the dipole response originates in a topological phase, it is not itself ``topological'' but depends sensitively on the geometry used to probe it.

\section{
Topological field theories for space-time dependent responses
in topological insulators and superconductors
in general dimensions from anomalies
}

\label{ChapterAnomalies}

The previous sections of this paper complete the list of the (topological) field theories
describing the space-time dependent responses of all topological insulators and superconductors
in three spatial dimensions ($3+1$ space-time dimensions). 
In this section we will describe, more generally, the (topological) field theories for
such responses in general dimensions. 
Most importantly, the main result obtained in this
section is a general connection between the appearance of
such 
%\textcolor{red}{
%%(topological) field theories 
%topological term in field theories }
topological terms in the field theories for the responses 
and 
the appearance of what are called 
anomalies\cite{WillExplainAnomalies}
for the field theories 
in those space-time dimensions in which topological insulators 
(superconductors) appear.
%\textcolor{red}{
%An anomaly, in particular when the bulk states are characterized 
%by an integer-valued topological invariant,
%is the breakdown of \textit{local} conservation law due to
%quantum effects, 
%which can be used to characterize the topological bulk. 
%}
In fact, we may 
ask if the existence of a particular type
of anomaly in a given dimension allows us to
predict
the existence of a topological
insulator (superconductor) of the `ten-fold' classification in that dimension.
The answer to this question is affirmative. As we demonstrate below,
a large part of the `ten-fold' classification can be derived from the existence
of the known anomalies
in corresponding quantum field theories in space-time. 
%\cite{footnoteInteger}
This can then be thought
of as yet another derivation of the `ten-fold' classification,
in addition to the previously known derivations such as that
based on Anderson localization at 
the sample boundaries\cite{SRFL},
 and
K-theory\cite{kitaevclassification} (as
well as a later point of view based on D-branes
\cite{DbranesTI,DbranesTIPRD}).
Moreover, and most importantly, the appearance of an anomaly 
is a statement about the respective quantum field 
theory (of space-time linear responses)
independent
 of the assumption of the absence of inter-particle
interactions. Thus, anomalies provide a description of
topological insulators (superconductors) in the context
of interacting systems.

\subsection{Topological insulators (superconductors) in the two complex symmetry classes A and AIII
from anomalies in the gauge field action}
\label{Topological insulators (superconductors) in the two complex symmetry classes A and AIII
from anomalies in the gauge field action}

\subsubsection{the integer quantum Hall effect (class A)}
\label{the integer quantum Hall effect (class A)1}

Let us begin by describing the topological field theories
describing the space-time dependent responses 
of the two ``complex'' symmetry classes, classes A and AIII in the Cartan (Altland-Zirnbauer)
classification.
\cite{SRFL,SRFLLandau100,SRFLnewJphys}
This includes the most familiar example, namely the integer quantum Hall insulator (IQH),
belonging to symmetry class A.
In both symmetry classes, A and AIII, there has to exist
a conserved $U(1)$ charge (particle number). This is the electromagnetic
charge, since these symmetry classes can be realized as normal electronic systems
(as opposed to superconducting quasiparticle systems) \cite{footnote2}.
Therefore we can minimally couple these topological insulators to an external
$U(1)$ gauge field. The field theory describing the space-time dependent linear responses
of the topological insulator can then be obtained by integrating out the gapped fermions.
The fact that the underlying insulator is topological is reflected in
the fact that the effective action for the external
$U(1)$ gauge field, describing 
the electromagnetic linear responses,
contains a term of
`topological origin', such as e.g. a Chern-Simons- or a $\theta$-term,
or corresponding higher dimensional analogues of these terms (see below
for more details).

In turn, the presence of terms of topological origin in the so-obtained effective
action for the external $U(1)$ gauge field are closely related
to the presence
 of a so-called anomaly. To see how an anomaly 
for the theory of the external $U(1)$ gauge field can actually predict
the presence of a topological phase, let us consider
first, as the simplest example,  the IQH insulator in $d=2$ spatial dimensions
-- symmetry class A.
(The space-time dimension is thus $D=2+1$).
In fact, let us first focus attention on the theory of the sample
boundary (the edge state), which has $d=1$ spatial dimensions. 
It is known (see below) 
that the effective theory for the linear responses of the $U(1)$ gauge
field in $D=1+1$ space-time dimensions (i.e. of the edge state)
can have what is called a ``gauge anomaly''
since the space time dimension $D$ is even.
\cite{nakahara, fujikawa}
The presence of this anomaly simply means that $U(1)$ charge conservation
is spoiled by quantum mechanics. In the condensed matter setting 
of the IQH insulator the meaning of this anomaly is that the
system (i.e. the edge) in $D=1+1$ space-time dimensions,
exhibiting the anomaly, does not exist in isolation, but is
necessarily realized as the boundary of a topological insulator in
one dimension higher. In this case, the breakdown of the conservation law
of $U(1)$ charge conservation at the boundary simply means that the current
``leaks'' into the bulk. Thus, in the condensed matter setting, the presence
of the anomaly in the theory at the boundary is not something abnormal,
but it is a physical effect: it is the integer quantum Hall effect. As we will
discus shortly below, the same reasoning applies to all even
space-time dimension, $D=2k$. Consequently, we see that the presence
of a $U(1)$ gauge anomaly predicts the presence of a topological insulator
in one dimension higher. 
I.e., this predicts
the presence of a topological
insulator in symmetry class A in $D=2k+1$ space-time dimensions,
in agreement with the `ten-fold' way classification.

\subsubsection{three-dimensional insulator (superconductor) in symmetry class AIII}
\label{three-dimensional insulator (superconductor) in symmetry class AIII}

Let us now consider the topological insulator (superconductor) in the
other complex symmetry class, class AIII, in $d=3$ spatial dimensions.
Again, the space-dimension $D=3+1=4$ is even.
It is known (see below) that in all even space-time dimensions the
effective action for the space-time dependent
$U(1)$ gauge field may also possess 
a different anomaly [in contrast to
the discussion in 
subsection \label{the integer quantum Hall effect (class A)} above],
often referred to as the 
``chiral (or: axial) anomaly in a background $U(1)$ gauge field''. 
\cite{nakahara}
The meaning of such an anomaly can be explained using Eq.\ (\ref{ChiralAnomalyInBackgroundGaugeField})
below: the so-called axial (or: chiral) $U(1)$ current $J^\mu_5(x)$
is {\it not} conserved in the presence of a background
$U(1)$ gauge field, i.e. $D_\mu J^\mu_5(x)\not =0$,
where $D_\mu$ denotes the covariant derivative
in the presence of a background gauge field. In the
simplest case of a single copy of a massive Dirac fermion (mass $m$),
%this derivative is given by Eq.\ (\ref{ChiralAnomalyInBackgroundGaugeField}).
this covariant derivative of the current 
is given by Eq.\ (\ref{ChiralAnomalyInBackgroundGaugeField}) below.
As displayed in this equation, there
are two sources of the lack of conservation: (i) a finite mass $m\not =0$
and (ii) the extra ``anomaly'' term 
${\cal A}_{2n+2}$
(to be discussed in more detail below),
which represents the breaking of the conservation of $J^\mu_5$
by quantum effects.
\cite{footnote3}
Now, as discussed in Ref.\ \onlinecite{Pavan09}, 
the presence of a 
``chiral (or: axial) anomaly in a background $U(1)$ gauge field'' 
implies directly the possibility of having a non-vanishing
$\theta$-term when deriving the effective action for
the external $U(1)$ gauge field
\cite{footnote4}.
(The $\theta$-angle is fixed\cite{SRFLnewJphys} to $\theta=\pi$ by a discrete
symmetry, which is the chiral symmetry for symmetry class AIII.)
Thus, the presence of 
a ``chiral (or: axial) anomaly in a background $U(1)$ gauge field'' 
in $D=2k$ space-time dimensions
signals
the existence of a topological insulator in this
space-time dimension through the appearance of
a $\theta$-term in the (topological) field theory for
the linear responses.

\subsubsection{anomaly polynomials and descent relation}
\label{anomaly polynomials and descent relation}

Observe that above we have used  anomalies of {\it two kinds},
and we used them  {\it in two different ways}:

\noindent (i): In case 1 there was an anomaly in the theory 
of the responses at the
{\it boundary} [which had $D=(d-1)+1$ space-time dimensions].
In this case the anomalous theory (i.e. the one at the boundary)
was gapless (critical); we refer to this situation as
a gauge anomaly (i.e., non-conservation of the $U(1)$ charge in question).
The presence of this anomaly implied the existence of a topological
insulator in one dimension higher, i.e. in $D'=d+1$ space-time dimensions.
The responses of this 
topological insulator are described by an effective Chern-Simons
action for the $U(1)$ gauge field in $D'=d+1$ space-time dimensions.
[See also Eq.\ (\ref{DescentRelation})].

\noindent (ii): In case 2 there existed an anomaly in the massive {\it bulk}
theory  in $D=d+1$ space-time dimensions. This was a chiral anomaly
(referring to the violation of the 
conservation of the global {\it axial}
$U(1)$ current $J_5^\mu$) in
the background of a non-vanishing $U(1)$ background
gauge field.

There are important relationships between
the following
different anomalies,
\begin{itemize}
\item
the $U(1)$ gauge anomaly in $D=2n$,
\item
the Chern-Simons term (i.e., parity anomaly) in $D=2n+1$ and
\item 
chiral anomaly in the presence of a background gauge field in $D=2n+2$,
\end{itemize}
which can be summarized, in terms 
of the so-called {\it descent relation}  of the ``anomaly polynomial''
\cite{nakahara}. 
Let us now explain this relation.

As mentioned above, it is known that
in even spacetime dimensions $D=2n$, there is a $U(1)$ gauge anomaly. 
If there is a gauge anomaly, the (Euclidean) effective action $\ln Z[\mathcal{A}]$ 
in the presence of the gauge field $\mathcal{A}$ 
is not invariant under a gauge transformation 
$\mathcal{A}\to \mathcal{A}+v$. Thus we can write
\begin{eqnarray}
\delta_v \ln Z[\mathcal{A}] = 2\pi {i} \int_{M_{2n}}
\Omega^{(1)}_{2n}(v, \mathcal{A},\mathcal{F}). 
\end{eqnarray}
where the variation $\delta_v$ is the gauge transformation in question,
and $\Omega^{(1)}_{2n}$ is a $2n$-form 
built from 
the connection 1-form, 
$\mathcal{A}=A_{\mu}dx^{\mu}$,
its field-strength 2-form,
$\mathcal{F}=(1/2)F_{\mu\nu}dx^{\mu}dx^{\nu}$,
and the variation $v=v_\mu dx^{\mu}$ of the gauge field.
(By definition, $\Omega^{(1)}_{2n}$ is linear in $v$.
The integral is taken over the physical $D=2n$ dimensional (Euclidean) space-time
${M_{2n}}$.)
Now, the descent relation tells us that
$\Omega^{(1)}_{2n}$ can be derived from 
the so-called anomaly polynomial $\Omega_{2n+2}(\mathcal{F})$,
which is a $2n+2$-form built from the curvature 2-form $\mathcal{F}$,
with the aid of yet another $2n+1$-form $\Omega^{(0)}_{2n+1}$, 
by 
\begin{eqnarray}
\label{DescentRelation}
\Omega_{2n+2} = d\Omega^{(0)}_{2n+1}, 
\quad 
\delta_v \Omega^{(0)}_{2n+1} = d\Omega^{(1)}_{2n}. 
\end{eqnarray}
I.e., $\Omega_{2n+2}$
is closed, 
and gauge invariant, and hence can be written 
as a polynomial in $\mathcal{F}$.
Here $\Omega^{(0)}_{2n+1}(\mathcal{A},\mathcal{F})$
is its corresponding Chern-Simons form.

\noindent There is a simple closed form expression for
the anomaly polynomial $\Omega_{2n+2}$ which is given by
\begin{eqnarray}
\label{DEFAnomalyPolynomial}
\Omega_{D}(\mathcal{F})
=
\mathrm{ch}(\mathcal{F})|_{D}
% =
% \lim_{M\to \infty}
% \frac{M^{2n}}{(4\pi)^n}
% \mathrm{tr}\, 
% \gamma_{2n+1}
% \exp\left[
% \frac{F}{M^2}
% \right]
% \mathrm{det}\,
% \left|
% \frac{ \mathrm{i} \hat{R}/(2M^2)}
% {
% \sinh (\mathrm{i} \hat{R}/(2M^2) )
% }
% \right|^{1/2}
\end{eqnarray}
Let us explain the notation: 
$\mathrm{ch}(\mathcal{F})$ is 
the following power series ("characteristic class")
constructed from the 
field-strength two-form $\mathcal{F}$ 
and is given by
\begin{eqnarray}
\label{CharacteristicClass}
\mathrm{ch}(\mathcal{F})
=
r + \frac{{i}}{2\pi} \mathrm{tr}\,  \mathcal{F}
- \frac{1}{2(2\pi)^2} \mathrm{tr}\, \mathcal{F}^2
+\cdots.
\end{eqnarray}
This expression is written for the general case of a 
gauge field transforming in a $r$-dimensional
irreducible representation 
of a (possibly non-Abelian)
gauge group, where $\mathrm{tr}$ denotes the trace in this representation.
Observe that $\mathrm{ch}(\mathcal{F})$
consists of a sum of different
$p$-forms with different $p$
where $p=\mbox{even}$.
The
notation
$\cdots |_{D}$
in Eq.\ (\ref{DEFAnomalyPolynomial})
means we extract a $D$-form from $\mathrm{ch}(\mathcal{F})$.

While up to this point the differential forms $\Omega^{(0)}_{2n+1}$
and $\Omega_{2n+2}$  appear to have been introduced solely to express 
the $D=2n$-dimensional gauge anomaly
in terms of other objects, 
they themselves are known to be related to 
other types of anomalies: 
the Chern-Simons form 
$\Omega^{(0)}_{2n+1}$ represents 
an anomaly in a discrete symmetry
(parity or charge-conjugation symmetry, depending on dimensionality)
discussed in more detail in subsection \ref{the Chern-Simons term}, 
below, 
and
$\Omega_{2n+2}$ represents\cite{nakahara} the chiral anomaly in the presence of
a background gauge field, discussed in 
subsection 
\ref{three-dimensional insulator (superconductor) in symmetry class AIII}
above. 
The integral of $\Omega_{2n+2}$ over $D=2n+2$ dimensional
space-time, on the other hand,
represents the $\theta$ term 
(see also subsection \ref{the theta-term} below).

\subsubsection{the Chern-Simons term}
\label{the Chern-Simons term}
The integral of $\Omega^{(0)}_{2n+1}(\mathcal{A},\mathcal{F})$ 
over $D=2n+1$-dimensional space-time
is the Chern-Simon type action for the gauge field
$\mathcal{A}$, 
and represents, as already mentioned,
an anomaly in a discrete symmetry: the parity or charge-conjugation anomaly. 

In turn, the presence of such a Chern-Simons term in the effective (bulk) action
for the gauge field $\mathcal{A}$ in
$D=2n+1$-dimensional space-time
signals the presence of a topological phase:
when there is a boundary in the system, 
the integral of the Chern-Simons term
is not invariant on its own;
rather, upon
making use of the descent relation
Eq.\ (\ref{DescentRelation}), one obtains
\begin{eqnarray}
\label{AboutTheChernSimonsForm}
\delta_v \int_{M_{2n+1}} \Omega^{(0)}_{2n+1}
=
 \int_{M_{2n+1}} d\Omega^{(1)}_{2n}
=
\int_{\partial M_{2n+1}} \Omega^{(1)}_{2n}. 
\end{eqnarray}
This is something we are familiar with
from the physics of the quantum Hall effect:
the presence of the boundary term 
$\int_{\partial M_{2n+1}} \Omega^{(1)}_{2n}$
appearing on the
right hand side
of Eq.\ (\ref{AboutTheChernSimonsForm})
signals the presence of edge mode.
In turn, as we have seen in 
subsection \ref{the integer quantum Hall effect (class A)}, 
the gauge anomaly in $D=2n$ dimensional space-time
which is represented by the integral over
$\Omega^{(1)}_{2n}$, itself
signals the presence of a topological phase in $D=2n+1$
space-time dimensions, i.e. in one dimension higher.

\subsubsection{the theta-term}
\label{the theta-term}

The integral of the anomaly polynomial $\Omega_{2n+2}$
over $D=2n+2$ dimensional space-time
is the $\theta$-term
and represents a chiral anomaly in the presence of a background
gauge field (discussed in subsection 
\ref{three-dimensional insulator (superconductor) in symmetry class AIII} above).
Again, to be more explicit, in the presence of such an axial anomaly, 
the axial current $J_5^{\mu}(x)$ 
(which in the present case
is an axial $U(1)$ current) is not conserved: 
$D_{\mu} J_5^{\mu}(x)\neq 0$
where $D_{\mu}$ is the covariant derivative in the presence of
the gauge field. 
For a single copy of a massive Dirac fermion, it is given by
\begin{eqnarray}
\label{ChiralAnomalyInBackgroundGaugeField}
D_{\mu} J_5^{\mu}(x)
=
2 {i} m \bar{\psi} \gamma_{2n+1} \psi
+
2 {i} \mathcal{A}_{2n+2} (x),
\end{eqnarray}
where the first term represents the explicit breaking 
of the chiral symmetry by the mass term,
whereas the 2nd term represents the breaking of the chiral 
symmetry by quantum effects.
$\mathcal{A}_{2n+2}$ quantifying 
the breaking of the axial current conservation by an anomaly
is essentially identical to $\Omega_{2n+2}$,
and given by removing all $dx^{\mu}$
which appear in the differential form $\Omega_{2n+2}$. 

Just as it was the case for the Chern-Simons term,
the presence of such a $\theta$-term in the effective action
for the gauge field
signals the presence of a topological phase. In particular,
the descent relation tells us that
\begin{eqnarray}
\int_{M_{2n+2}} \Omega_{2n+2}
=
 \int_{M_{2n+2}} d\Omega^{(0)}_{2n+1}
=
\int_{\partial M_{2n+2}} \Omega^{(0)}_{2n+1}
\nonumber \\%%%%%
\end{eqnarray}
This is, again, something we are familiar with from the physics of 
the three-dimensional topological insulator in class AIII,
which is described by the $\theta$-term (the axion-term).
In the presence of  a boundary $\partial M_{2n+2}$, such a topological state 
supports boundary degrees of freedom, as signaled by
the boundary term
$\int_{\partial M_{2n+1}} \Omega^{(0)}_{2n+1}$
which is a Chern-Simons term.
\cite{footnote5}

Let us summarize:
in order to derive the existence of topological phases in symmetry class A and AIII,
we start from the anomaly polynomial $\Omega_{2n+2}$. 
Then the terms 
$\int_{M_{2n+2}}\Omega_{2n+2}$
and
$\int_{M_{2n+1}}\Omega^{(0)}_{2n+1}$
are
the effective actions for the (topological)
field theory of the space-time linear responses for the gauge field
for the  topological phases in class AIII ($D=2n+2$) and A ($D=2n+1$), respectively.

\subsection{Topological insulators (superconductors) in the remaining eight
`real' symmetry classes from gravitational and mixed anomalies}

\subsubsection{Gravitational anomaly and axial anomaly in the presence
of background gravity}

For the remaining eight ``real'' of the ten  symmetry classes, 
having a conserved $U(1)$ quantity is less trivial.
Classes AI,  AII, and CII
are naturally realized as a normal (as opposed to superconducting)
electronic system,
and thus for these there is a natural notion of a conserved $U(1)$ quantity
(the electrical charge).  One realization of the BDI symmetry class,
which is only part\cite{footnote-Explain-BDI} 
of the entire symmetry class, can also
be considered to have a conserved  $U(1)$ 
quantity and we consider this realization in this subsection.
On the other hand, classes  D, DIII, C and CI
are naturally realized as  BdG systems.
While for classes C and CI, $SU(2)$ spin is conserved (so a conserved $U(1)$ charge
exists), 
for classes D and DIII, there is no conserved $U(1)$ quantity at all.

Since for the latter four  of eight real symmetry classes
(D, DIII, C, CI)
we cannot rely on a conserved $U(1)$ quantity to describe these 
topological phases, it is not possible to couple these systems
minimally to a $U(1)$ gauge field. However, 
it is natural to consider a coupling of  these topological phases 
to gravity. 
Let us focus first 
on topological insulators (superconductors)
with a integer topological charge, $\mathbb{Z}$,
but not on those with a binary topological charge, $\mathbb{Z}_2$. 
For now we
also do not consider topological insulators/superconductors with a $2\mathbb{Z}$ charge.

An analogue of the $U(1)$ gauge anomaly,
which we have described in 
section \ref{the integer quantum Hall effect (class A)}
at the boundary 
(of space-time dimension $D=2n$)
of topological phases 
in symmetry class A
is the gravitational anomaly. 
It corresponds to
the breakdown of energy-momentum conservation,
and when it happens, it must be realized in a system which represents
the boundary of a topological phase in one dimension higher
(in analogy to the case of a $U(1)$ gauge anomaly, 
section \ref{the integer quantum Hall effect (class A)}).
We refer to this anomaly also as a `purely gravitational anomaly'.
In the following we will show  that one can
predict the appearance of the topological phases
in symmetry classes D, C, DIII, CI 
[i.e. those without conserved $U(1)$ charge]
from
the presence of a purely gravitational anomaly which
appears in the field theory 
for the gravitational
(or: thermal\cite{Luttinger1964}) responses.

Finally, we will need to discuss the still remaining symmetry classes
AI, BDI, AII, and CII.
Topological insulators (superconductors) in these
symmetry classes can be coupled to both, a $U(1)$ gauge
field\cite{RecallCommentAboutBDI}
as well as a gravitational background.
We will show that
the field theories for the  space-time dependent linear responses 
for these topological insulators
possess a so-called mixed anomaly.
Indeed, we will show that
the appearance of a mixed gravitational and electromagnetic 
axial anomaly signals the existence of topological phases in these
symmetry classes.

\subsubsection{Topological insulators (superconductors)
in symmetry classes D, C, DIII, CI, from
the purely gravitational anomaly}

As mentioned earlier in this paper, each topological
insulator (in any dimension) has a Dirac Hamiltonian
representative. 
\cite{SRFLnewJphys}
We can consider the coupling
of 
this Dirac theory
to a space-time dependent gravitational
background. Upon integrating out the massive
fermions, we obtain an effective gravitational
action in $D$ space-time dimensions.
If there is a gravitational anomaly, the (Euclidean) 
effective action $\ln Z[e,\omega]$ 
in the presence of the gravitational background
is not invariant under a general coordinate transformation 
$x^{\mu}\to x^{\mu}+\epsilon^{\mu}$,
where $e$ is the  vielbein and $\omega$ is the spin-connection one-form.
I.e.,
\begin{eqnarray}
\delta_v \ln Z[e,\omega] = 2\pi {i} \int_{M_{D}}\Omega^{(1)}_{D}(v,\omega,\mathcal{R}), 
\end{eqnarray}
where $\delta_v$ represents an infinitesimal $SO(D)$ rotation,
under which $\omega$, the spin-connection 1-form $\omega$, is transformed as
$\omega\to \omega+v$;
$\Omega^{(1)}_{D}(v,\omega,\mathcal{R})$ is a $D$-form 
related to the gravitational anomaly.
In complete analogy 
to the case of the gauge anomaly discussed above,
$\Omega^{(1)}_{D}(v, \omega,\mathcal{R})$
can be derived from 
a corresponding 
anomaly polynomial 
$\Omega^{}_{D+2}(\mathcal{R})$
(see Eqs.\ (\ref{GravitationalAnomalyPolynomial}), (\ref{DiracGenus}) below)
through its Chern-Simons form
$\Omega^{(0)}_{D+1}(\omega, \mathcal{R})$,
by using 
a descent relation 
which
takes a form identical
to
Eq.\ (\ref{DescentRelation}).
Thus, once the existence of the (purely) gravitational anomaly is known 
for a given dimension $D$, it predicts the presence of topological 
phases in $D+1$ and $D+2$ dimensions, using the same logic
as in the gauge field case above.

Now, according
to Ref.\ \onlinecite{AlvarezGaumeWitten83}, 
a purely gravitational anomaly can exist in 
\begin{eqnarray}
D= 4k+2
\quad
(d=4k+1). 
\end{eqnarray}
Thus, 
breakdown of energy-momentum conservation 
due to quantum effects
can occur 
in these dimensions. 
As in the case of symmetry class A, discussed above,
we take this as evidence for the existence of a topological bulk 
in one dimension higher, i.e. in space-time dimensions
\begin{eqnarray}
D= 4k+3
\quad
(d=4k+2). 
\end{eqnarray}
This thus predicts
the appearance of
topological phases in 
\begin{eqnarray}
\mbox{class D ($d=2$)},
\quad
\mbox{class C ($d=6$)},
\end{eqnarray}
as well as all the other higher-dimensional topological phases
that we can obtain from these by Bott periodicity. 
(These are colored red in 
Table \ref{periodic table with gravity}.)

On the other hand, there is an analog of the
``axial anomaly in the presence of a background gauge field'' 
which we discussed in section 
\ref{three-dimensional insulator (superconductor) in symmetry class AIII}
in the context of symmetry class AIII in $D=2n$
space-time dimensions.
This analog  is the ``axial anomaly in the presence
of a background gravitational
field''. If only a background gravitational
field is present,
this anomaly
exists in space-time dimensions
\begin{eqnarray}
D= 4k
\quad
(d=4k-1). 
\end{eqnarray}
This covers symmetry classes
\begin{eqnarray}
\mbox{class DIII ($d=3$)},
\quad
\mbox{class CI ($d=7$)},
\end{eqnarray}
as well as
 all higher-dimensional topological phases
that we can obtain from these by Bott periodicity. 
(These are colored blue in Table \ref{periodic table with gravity}.)

The anomaly polynomial related to 
the gravitational
anomalies is known explicitly. It can be written as
\begin{eqnarray}
\label{GravitationalAnomalyPolynomial}
\Omega_{D=4k}
\!\!&=&\!\!
\hat{A}(\mathcal{R})|_{D} 
\end{eqnarray}
where 
$\hat{A}(\mathcal{R})$ is the 
so-called
Dirac genus given by\cite{fujikawa}
\begin{eqnarray}
\label{DiracGenus}
\hat{A}(\mathcal{R})
&=&
1 
+ \frac{1}{(4\pi)^2}\frac{1}{12} \mathrm{tr}\, \mathcal{R}^2 
\nonumber \\%%%%%
&&
+ \frac{1}{(4\pi)^2}
\left[
\frac{1}{288} (\mathrm{tr}\, \mathcal{R}^2 )^2
+
\frac{1}{360} \mathrm{tr}\, \mathcal{R}^4
\right]
+
\cdots.
\label{dirac genus}
\end{eqnarray}
%\textcolor{red}{(a typo is corrected here 
%according to the referee report).}
Here 
$\mathcal{R}$ is the $D\times D$ matrix of two-forms
\begin{equation}
\label{DEFcalR}
{\mathcal{R}_\mu}^\nu
:=
{1\over 2} {R_{\alpha\beta\mu}}^\nu \ dx^\alpha dx^\beta
\end{equation}
where $ {R_{\alpha\beta\mu}}^\nu$ is the usual  Riemann curvature
tensor, and the trace refers to the $D\times D$ matrix structure.
This defines, by the descent relation 
[which takes
a form identical to Eq.\ (\ref{DescentRelation})], 
the differential forms
$\Omega^{(0)}_{4k-1}$ and $\Omega^{(1)}_{4k-2}$. 
As before,
the notation
$\hat{A}(\mathcal{R})|_{D}$ extracts a $D$-form 
from $\hat{A}(\mathcal{R})$. 
It is obvious from (\ref{dirac genus}),
that the anomaly polynomial exists only for $D=4k$
because Eq.\ (\ref{dirac genus}) is a function of ${\cal R}^2$.
(Note that the descent relation Eq.\ (\ref{DescentRelation})
then implies the existence
of a purely gravitational anomaly
$\Omega^{(1)}_{4k+2}({\cal R})$ in $D=
4k+2$ space-time dimensions, in agreement with
Ref.\ \onlinecite{AlvarezGaumeWitten83}.)

\begin{table}
\begin{center}
\begin{tabular}{cccccccccccccc}\hline
Cartan$\backslash d$ & 
\textcolor{magenta}{0}  & 
\textcolor{green}{1} & 
\textcolor{red}{2} & 
\textcolor{blue}{3} & 
\textcolor{magenta}{4} & 
\textcolor{green}{5} & 
\textcolor{red}{6} & 
\textcolor{blue}{7} & 
\textcolor{magenta}{8} & 
\textcolor{green}{9} & 
\textcolor{red}{10} & 
\textcolor{blue}{11} & $\cdots$ \\ \hline\hline\hline
A   & $\mathbb{Z}$& 0  & $\mathbb{Z}$& 0 &  $\mathbb{Z}$& 0  & $\mathbb{Z}$& 0 & 
 $\mathbb{Z}$& 0 &  $\mathbb{Z}$& 0 &  $\cdots$ \\ \hline
AIII & 0& $\mathbb{Z}$& 0  & $\mathbb{Z}$& 0 &  $\mathbb{Z}$& 0  & $\mathbb{Z}$& 0 & 
 $\mathbb{Z}$& 0 &  $\mathbb{Z}$&  $\cdots$ \\ \hline\hline
AI  & \textcolor{magenta}{$\mathbb{Z}^{\spadesuit}$} & 0 & 0 
    & 0 & $2\mathbb{Z}$ & 0 
    & $\mathbb{Z}_2$ & $\mathbb{Z}_2$ & 
\textcolor{magenta}{$\mathbb{Z}^{\spadesuit}$}
    & 0 & 0 & 0 & $\cdots$ \\ \hline
BDI & $\mathbb{Z}_2$ & 
\textcolor{green}{$\mathbb{Z}^{\clubsuit}$} & 0 & 0 
    & 0 & $2\mathbb{Z}$ & 0 
    & $\mathbb{Z}_2$ & $\mathbb{Z}_2$ & 
\textcolor{green}{$\mathbb{Z}^{\clubsuit}$}
    & 0 & 0 &  $\cdots$ \\ \hline
D   & $\mathbb{Z}_2$ & $\mathbb{Z}_2$ & 
\textcolor{red}{$\mathbb{Z}^{\heartsuit}$}
    & 0 & 0 & 0 
    & $2\mathbb{Z}$  & 0 & $\mathbb{Z}_2$ 
    & $\mathbb{Z}_2$ & 
\textcolor{red}{$\mathbb{Z}^{\heartsuit}$}  & 0 & $\cdots$ \\ \hline
DIII& 0 & $\mathbb{Z}_2$ & $\mathbb{Z}_2$ & 
\textcolor{blue}{$\mathbb{Z}^{\diamondsuit}$ }
    & 0 & 0 & 0 
    & $2\mathbb{Z}$  & 0 & $\mathbb{Z}_2$ 
    & $\mathbb{Z}_2$ & 
\textcolor{blue}{$\mathbb{Z}^{\diamondsuit}$}  &  $\cdots$ \\ \hline
AII & $2\mathbb{Z}$  & 0 & $\mathbb{Z}_2$ 
    & $\mathbb{Z}_2$ & \textcolor{magenta}{$\mathbb{Z}^{\spadesuit}$} & 0 
    & 0 & 0 & $2\mathbb{Z}$ 
    & 0 & $\mathbb{Z}_2$ & $\mathbb{Z}_2$&  $\cdots$\\ \hline
CII & 0 & $2\mathbb{Z}$  & 0 & $\mathbb{Z}_2$ 
    & $\mathbb{Z}_2$ & 
\textcolor{green}{$\mathbb{Z}^{\clubsuit}$} & 0 
    & 0 & 0 & $2\mathbb{Z}$ 
    & 0 & $\mathbb{Z}_2$ &  $\cdots$\\ \hline
C   & 0  & 0 & $2\mathbb{Z}$  
    & 0 & $\mathbb{Z}_2$  & $\mathbb{Z}_2$ 
    & \textcolor{red}{$\mathbb{Z}^{\heartsuit}$} & 0 & 0 
    & 0 & $2\mathbb{Z}$ & 0 & $\cdots$ \\ \hline
CI  & 0 & 0  & 0 & $2\mathbb{Z}$  
    & 0 & $\mathbb{Z}_2$  & $\mathbb{Z}_2$ 
    & \textcolor{blue}{$\mathbb{Z}^{\diamondsuit}$} & 0 & 0 
    & 0 & $2\mathbb{Z}$  & $\cdots$ \\ \hline
\end{tabular}
\end{center}
\caption{
Topological insulators (superconductors) with an
integer ($\mathbb{Z}$) classification,  (A): in
the complex symmetry classes, predicted from the
chiral $U(1)$ anomaly, and (B):
%Integer topological phases 
in the real symmetry classes,
predicted from
the gravitational anomaly (red, ${\cdots}^{\heartsuit}$),
the chiral anomaly in the presence of background gravity 
(blue, ${\cdots}^{\diamondsuit}$),
and the chiral anomaly in the presence of both background gravity and 
$U(1)$ gauge field 
(green, ${\cdots}^{\clubsuit}$).
}
\label{periodic table with gravity}
\end{table}

\subsubsection{Topological insulators (superconductors)
in symmetry classes AI, BDI, AII, CII from the mixed anomaly}

Before proceeding let us briefly summarize the previous subsection:
by considering various anomalies related to gravity,
we can predict the integer topological phases in the BdG symmetry classes
D, DIII, C, and CI.
(As mentioned above, for a moment, we do not consider 
topological phases with $\mathbb{Z}_2$ or $2\mathbb{Z}$ topological
charges). 
On the other hand,
we have so far not covered
the description of topological insulators
in symmetry classes
AI, BDI, AII, and CII
in terms of anomalies.

So far, we have considered for the `real' symmetry classes
only those anomalies which involve 
solely
gravity. 
Since 
the (gapped) topological insulators in symmetry
classes
AI, BDI, AII, and CII, also possess a conserved $U(1)$ 
charge\cite{RecallCommentAboutBDI},
we can couple those to both, a $U(1)$ gauge field 
as well as a gravitational background.
Therefore, it is natural to consider an anomaly
which occurs in the presence of both,  a background gauge and
a background gravitational field.

As it turns out, even
in the presence of
both gauge and gravitational fields, 
the structure of the anomaly is similar to the one discussed so far:
the non-invariance of the effective action 
under a gauge transformation or coordinate transformation can be 
expressed as
\begin{eqnarray}
\delta_v \ln Z[\mathcal{A}, e,\omega] = 2\pi {i} \int_{M_{D}}
\Omega^{(1)}_{D}(v, \mathcal{A}, \omega, \mathcal{F},\mathcal{R}),
\end{eqnarray}
where $\Omega^{(1)}_{D}(v, \mathcal{A}, \omega, \mathcal{F},\mathcal{R})$ can be derived from 
an associated anomaly polynomial which reads,
\cite{fujikawa,nakahara}
\begin{eqnarray}
\label{MixedAnomalyPolynomialGeneratingFunction}
\Omega_{D} (\mathcal{R},\mathcal{F})
=
\left ( \mathrm{ch}(\mathcal{F}) \hat{A}(\mathcal{R}) \right )|_{D}.
\end{eqnarray}
As the right hand side is given simply given by the product of
the anomaly polynomials for a gauge field 
[Eq.\ (\ref{CharacteristicClass})]
and gravity 
[Eq.\ (\ref{DiracGenus})], 
by switching off either $\mathcal{R}$ or $\mathcal{F}$,
we recover the results discussed in the previous subsections:
for all even spacetime dimensions
$D=d+1=2k$ ($k=1,2,\ldots$)
we obtain a non-vanishing anomaly polynomial
$\Omega_{D} (\mathcal{R}=0,\mathcal{F})=\Omega_{D}(\mathcal{F})$,
which we have already used to predict topological insulators/superconductors
in class A ($D=2k+1$) and AIII ($D=2k$).
For space-time dimensions $D=d+1=4k$ ($k=1,2,\ldots$)
we obtain a non-vanishing anomaly polynomial
$\Omega_{D} (\mathcal{R},\mathcal{F}=0)=\Omega_{D}(\mathcal{R})$,
which we have already used to predict topological insulators/superconductors
in class DIII ($D=4+8k$) and CI ($D=8+8k$).

On the other hand, 
while 
the anomaly polynomial 
$\Omega_D({\cal R}, {\cal F}=0) = \Omega_D({\cal R})$ vanishes
in $D=4k+2$ dimensions, 
the one obtained from 
Eq.\ (\ref{MixedAnomalyPolynomialGeneratingFunction}),
namely
$\Omega_{D} (\mathcal{R},\mathcal{F})$, 
is non-vanishing in these dimensions.

As before, the anomaly polynomial itself is related to a 
`chiral anomaly in the presence of both gauge field and gravity' 
of the massive bulk system in $D=4k+2$ space-time dimensions,
$
D_{\mu} J_5^{\mu}(x)
=
2 {i} m \bar{\psi} \gamma_{D-1} \psi
+
2 {i} \mathcal{A}_{D} (x)
$,
where 
$\mathcal{A}_{D} (x)$ is given in terms of 
$\Omega_{D} (\mathcal{R},\mathcal{F})$.
For this reason, one predicts 
%also a topological insulator (superconductor)
an additional topological insulator (superconductor)
in these space-time dimensions
(besides the one of Sec.\
\ref{three-dimensional insulator (superconductor) in symmetry class AIII}).  
Therefore, one predicts 
the occurrence of
topological phases in spatial dimensions $d=9, \ (d=1)$ and $d=5$, 
\begin{eqnarray}
\mbox{class BDI ($d=9, \ (d=1)$)},
\quad
\mbox{class CII ($d=5$)},
\end{eqnarray}
as well as of all higher-dimensional topological phases
that we can obtain from these by Bott periodicity.\cite{footnote-Explain-BDI-Low-D}
(These are colored green in
Table \ref{periodic table with gravity}).
Indeed, for class BDI and CII, we can realize 
these symmetry classes as a normal (i.e. not superconducting) system,
and hence they have a natural $U(1)$ charge\cite{RecallCommentAboutBDI}. 
The effective topological field theory for
the space-time dependent linear [electrical and gravitational (thermal)]
responses possesses a term of topological  origin of the form
$\int \Omega_{D} (\mathcal{R},\mathcal{F})$,
where $D=4k+2$.

Moreover, it turns out that a descent relation
which is identical in form to
Eq.\ (\ref{DescentRelation}) also holds
for the `mixed' anomaly polynomial defined in 
Eq.\ (\ref{MixedAnomalyPolynomialGeneratingFunction}).
Therefore, the space-time integral of the Chern-Simons form 
$\Omega^{(0)}_{4k+1}$ of
$\Omega_{4k+2}$, 
which is obtained from $\Omega_{4k+2}$
by using the descent relation,
$d\Omega^{(0)}_{4k+1} =
\Omega_{4k+2}$,
describes the term of topological origin
in the effective action for the linear
responses in $D=4k+1$ space-time  dimensions.
This corresponds to a ``mixed anomaly''
$\Omega^{(1)}_{4k}$ in the corresponding
boundary theory in $4k$ space-time dimensions.
For this reason, one predicts
the occurrence of additional topological insulators 
in spatial dimensionalities $d=0$ and $d=4$
(besides the ones in Sec.
\ref{the integer quantum Hall effect (class A)1}),
for the two symmetry classes
\begin{eqnarray}
\mbox{class AI ($d=0$)},
\quad
\mbox{class AII ($d=4$)},
\end{eqnarray}
as well as for all their higher dimensional equivalents obtained from the Bott periodicity
(These are colored magenta in Table \ref{periodic table with gravity}).

\subsubsection{Atiyah-Singer Index Theorem}

For all the symmetry classes with {\it chiral symmetry}
the hamiltonian can be brought into block off-diagonal form\cite{SRFL}. 
Above we have discussed all symmetry classes of
this form which possess topological insulators with a $\mathbb{Z}$
classification 
(i.e.  AIII in  $D=2n$,
 DIII in $D=4+8k$,  CI in $D=8+8k$,  CII in $D=6+8k$,  BDI in $D=10+8k$).
A Dirac hamiltonian $\mathcal{H}$ with chiral symmetry
possesses an index, and the Atiyah-Singer Index Theorem\cite{nakahara} 
relates the integral of the anomaly polynomial discussed above to this
index through the formula
\begin{equation}
\label{FormulaAtiyahSingerIndexTheorem}
{\rm index}(\mathcal{H})
=
\int_{M_{D}} \ \Omega_{D}({\cal R}, {\cal F})
\end{equation}
where
$\Omega_{D}({\cal R}, {\cal F})$
is the most general anomaly polynomial,
as defined in Eq.\ (\ref{MixedAnomalyPolynomialGeneratingFunction})
above.
Here, the Dirac hamiltonian $\mathcal{H}$ refers to the hamiltonian
in a gravitational background and a background (Abelian or
non-Abelian) gauge field.
The index
${\rm index}(\mathcal{H})$ is by definition an integer.
We note that it is
because of this theorem that the space time integral
of the anomaly polynomial represents a $\theta$ term
for the theory of the space time dependent linear
gauge and gravitational responses,
and that the $\theta$ terms only occur  
for symmetry classes possessing a chiral symmetry.

\subsubsection{Global gravitational anomalies}

The discussion that we have given so far
for the connection between anomalies and
topological insulators and superconductors
in ``the primary series'' (those located in the diagonal
of the periodic table and 
characterized by an integer topological invariant)
can be extended to 
some of the ``first and second descendants''
(the topological insulators and superconductors
in the same symmetry class, but in one and two dimensions 
less than the one with a $\mathbb{Z}$ invariant; these are
each characterized by a $\mathbb{Z}_2$ invariant).
We propose that for these we need to use so-called global anomalies,
instead of the so-called perturbative anomalies that we
have made use of in this section. 
Such anomalies do not affect infinitesimal, but large (of order one)
symmetry transformations.

It was found in Ref.\ \onlinecite{AlvarezGaumeWitten83}
that global gravitational anomalies can exist, given
certain assumptions are satisfied, 
(i): in $D=8k$,  in (ii):  $D=8k+1$ and (iii): in $D=4k+2$ space time dimensions.
If so, then  following
the same reasoning as above, the presence
of these anomalies would indicate the existence
of a topological insulator in one dimension higher (of
which the anomalous system is the boundary).
This would then indicate the existence of topological
insulators (superconductors) 
in space time dimensions
(i): $D=8k+1$,  (ii):  $D=8k+2$ and (iii): $D=4k+3$
[corresponding to spatial dimensions
(i): $d=8k$,  (ii):  $d=8k+1$ and (iii): $d=4k+2$].
Indeed, there exist 
$\mathbb{Z}_2$ topological insulators in these dimensions
(Table \ref{periodic table with gravity}).
More precisely, there exist {\it two} 
$\mathbb{Z}_2$ topological insulators in these dimensions
and at this point we have not yet explored in detail
which of the two (or if both) could be  related to this global gravitational anomaly.
Moreover we note that there also exist other (i.e. not gravitational)
global anomalies, and we propose that the other, so far not yet covered
$\mathbb{Z}_2$ topological insulators can be obtained from considering these other global
anomalies.

We end by mentioning that the
notions presented
in this Section (Section \ref{ChapterAnomalies})
may also be further supported 
by the connection with the tenfold classification 
of D-branes\cite{DbranesTI, DbranesTIPRD}:
In the D-brane realizations of topological insulators and superconductors, 
massive fermion spectra arise as 
%an 
open string excitations connecting two D-branes,
which are in one-to-one correspondence with
the Dirac representative of the ten-fold classification of topological insulators and superconductors, 
and come quite naturally with gauge interactions.
The Wess-Zumino term of the D-branes gives rise to a gauge field theory
of topological nature,
such as ones with the Chern-Simons term or the $\theta$-term
in various dimensions.

\section{Conclusions}
\label{Conclusions}

There are various important future research directions in the field
of topological insulators and superconductors. Let us mention two
here. One is the search for experimental realizations of the topological
singlet and triplet superconductors in three spatial dimensions,
besides the B phase of
the  $^{3}$He superfluid. 
Given how fast experimental realizations of the quantum
spin Hall effect in two spatial dimensions 
and the $\mathbb{Z}_2$ topological insulators 
in three dimensions have been found,
one may perhaps anticipate a similar development
for these three-dimensional topological
superconducting phases.
Notably, Cu$_x$Bi$_2$Se$_3$, 
which arises from the familiar three-dimensional topological insulators
Bi$_2$Se$_3$, was found to be superconducting at 3.8 K.  \cite{Hor2009}
Subsequent theoretical work proposed that this superconducting phase 
should be a topological superconductor.  \cite{FuBerg2009}
The various linear responses discussed in this paper,
as summarized in Table \ref{table1},
may become helpful in the search for, and for identifying
various such topological phases.

Another important issue is to complete the study of the effect of interactions
for the symmetry classes so far not yet included in the discussion
given in Section \ref{ChapterAnomalies}. (These include, in general
dimensionalities the topological insulators (superconductors)
with a  $2\mathbb{Z}$  classification, 
as well as  the majority of those with a
$\mathbb{Z}_2$ classification.) Moreover, this includes
the case of symmetry class BDI in $d=1$ spatial dimension
(recall also Refs.\ \onlinecite{footnote-Explain-BDI,footnote-Explain-BDI-Low-D}),
discussed in the work
of Refs.\  \onlinecite{FidkowskiKitaev1,FidkowskiKitaev2,TurnerPollmannBerg}. 
Further important outstanding questions
concern possible
topological phases (besides superconductors)
which may arise from interactions rather than from band effects.
How can one describe ``fractional'' versions of the topological
insulators (superconductors), 
\cite{fractional} 
and how can one classify
bosonic systems such as  e.g. spin systems? 
\cite{Wen}
Clearly, in order to address any of these 
interaction-dominated issues one cannot rely on a topological
invariant defined in terms of
single-particle Bloch wavefunctions.
Rather, a definition of topological quantum states of 
matter in terms of responses to physical probes is necessary.
In this paper we have developed  
a description of this type
for all topological insulators in three spatial dimensions,
and for a significant part of the topological insulators
in general dimensions.
From a conceptual point of view the gravitational
responses are the most fundamental ones, in that
they apply to all topological insulators. 
Owing to Luttinger's derivation\cite{Luttinger1964}
of the thermal Kubo formula, these correspond physically
to thermal responses functions.

\vskip .6cm

\acknowledgements

We thank
Taylor Hughes, Charles Kane, 
Alexei Kitaev,
Shunji Matsuura,
Xiao-Liang Qi,
Tadashi Takayanagi, 
Ashvin Vishwanath, 
and
Shou-Cheng Zhang 
for useful discussions.
SR thanks Center for Condensed Matter Theory at University of
California, Berkeley for its support.  
JEM acknowledges support from NSF DMR-0804413.  This work was supported, in part, by the NSF under Grant No. DMR-0706140 (A.W.W.L.).

%\bibliography{draft_toporesponses-1}

\begin{thebibliography}{99}
\bibitem{footnote-ChernSimons-Anomaly}
The theory of the space-time dependent responses
at the surface is the Chern-Simons theory (see Section
\ref{ChapterAnomalies} for more details).
The charge, spin, and thermal surface
conductivities (in natural units)
are the coupling constants of
the Chern-Simons terms.
\bibitem{footnote_reimann}


We hereby collect our conventions for 
metric, vielbein, spin connection, etc.
We start from the metric and its inverse,
\begin{eqnarray}
g^{\mu\nu},
\quad
g_{\rho \sigma},
\quad 
\mbox{with} 
\quad 
g^{\mu\nu} g_{\nu\rho} =  \delta^{\mu}_{\rho}. 
\end{eqnarray}
The components of the Levi-Civita connection is
\begin{eqnarray}
\Gamma^{\mu}{ }_{\nu\rho}
\!\!&=&\!\!
\frac{1}{2}
g^{\mu \alpha}
\left(
\partial_\nu g_{\rho \alpha}
+
\partial_\rho g_{\nu \alpha}
-
\partial_\alpha g_{\nu \rho}
\right).
\end{eqnarray}
The vielbein 
$
%{ }^{\ }_{{a}} e^{\mu}
%=
%{ }^{\mu} e^{\ }_{{a}}
%=
 e^{\ }_{{a}}{ }^{\mu}
$
and
$
%{ }^{{a}}e^{\ }_{\mu} 
%=
%{ }^{\ }_{\mu} e^{{a}}
%=
e^{{a}}{ }^{\ }_{\mu}
$
diagonalizes metric, and is defined by
\begin{eqnarray}
g^{\ }_{\mu\nu}
e^{\ }_{{a}}{ }^{\mu}
e^{\ }_{{b}}{ }^{\nu}
=
\eta^{\ }_{{a} {b}},
%\quad
%\left(
%{ }^{\ }_{{a}} e^{\mu}
%{ }^{\ }_\mu g^{\ }_{\nu}
%{ }^{\nu}e_{{b}}
%=
%{}^{\ }_{{a}}\eta^{\ }_{{b}}
%\right),
\quad
\eta^{\ }_{{a}{b}}
e^{{a}} { }^{\ }_{\mu}
e^{{b} } { }^{\ }_{\nu}
=
g^{\ }_{\mu \nu}. 
%\nonumber \\%%%%%
%\eta^{\ }_{{a}{b}}
%e^{{a}} { }^{\ }_{\mu}
%e^{{b} } { }^{\ }_{\nu}
%=
%g^{\ }_{\mu \nu}. 
%\quad
%\left(
%{ }^{\ }_{\mu} e^{{a}}
%{ }^{\ }_{{a}} \eta^{\ }_{{b}}
%{ }^{{b}}e^{\ }_{\nu}
%=
%{}^{\ }_{\mu}g^{\ }_{\nu}
%\right).
\end{eqnarray}
Here, $\eta_{ab}$ is a flat (Minkowski) metric 
and 
we use Greek indices 
$\mu,\nu,\ldots$ for coordinates of the manifold,
and 
Roman indices 
${a}, {b},\ldots$
for the flat  coordinates
at some point $x_0$ of the manifold;
they are raised and lowered by 
$g_{\mu\nu}, g^{\mu\nu}$ and
$\eta_{ab}, \eta^{ab}$, 
respectively. 
% Also
% \begin{eqnarray}
% { }^{\ }_\mu e^{{a}}
% { }^{\ }_{{a}} e^\nu
% =
% { }^{\ }_\mu \delta^\nu,
% \quad
% e^{{a}}{ }^{\ }_\mu
% e^{\ }_{{a}}{ }^{\nu}
% =
% { }^{\ }_\mu \delta^\nu,
% \nonumber \\%%%%%
% { }^{\ }_{{a}} e^{\mu}
% { }^{\ }_{\mu} e^{{b}}
% =
% { }^{\ }_{{a}} \delta^{{b}},
% \quad
%  e^{\ }_{{a}}{ }^{\mu}
% e^{{b}}{ }^{\ }_{\mu} 
% =
% { }^{\ }_{{a}} \delta^{{b}},
% \end{eqnarray}
Since the vielbein $e^{{a}}{ }_{\mu}$
transforms as a covariant vector under
general coordinate transformation,
it is convenient to introduce a one form
\begin{eqnarray}
e^{{a}}
=
e^{{a}}{ }_{\mu} dx^{\mu}.
\end{eqnarray}
The spin connection is
\begin{eqnarray}
\omega_\mu{ }^{{a}}{ }^{\ }_{{b}}
=
e^{{a}}{ }^{\ }_\alpha
\left[
\partial^{\ }_\mu
e^{\ }_{{b}}{ }^\alpha
+
\Gamma^{\alpha}{ }^{\ }_{\mu \beta} 
e^{\ }_{{b}}{ }^\beta
\right]. 
%\quad
%\left(
%{ }^{{a}}
%\left(\omega_\mu\right)^{\ }_{{b}}
%=
%{ }^{{a}}e^{\ }_\alpha
%\left[
%\partial^{\ }_\mu
%{ }^\alpha e^{\ }_{{b}}
%+
%\Gamma^{\alpha}{ }^{\ }_{\mu \beta} 
%{ }^\beta e^{\ }_{{b}}
%\right]
%\right)
\end{eqnarray}
This can be written in terms of a covariant vector 
$e^{\ }_{{b}}{ }^\mu$,  
%for a fixed ${b}$, 
which is the ${b}$-th eigenvector of the metric,
by using the covariant derivative with respect to 
the Levi-Civita connection $\Gamma^{\mu}{ }_{\nu\rho}$ as 
\begin{eqnarray}
\omega_\mu { }^{{a}}{ }^{\ }_{{b}}
\!\!&=&\!\!
e^{{a}}{ }^{\ }_\alpha
\nabla^{\ }_\mu
e^{\ }_{{b}}{ }^\alpha.
\end{eqnarray}
We define the connection one-form by 
\begin{eqnarray}
\omega^{{a}}{ }^{\ }_{{b}}
=
\omega_\mu { }^{{a}}{ }^{\ }_{{b}} dx^{\mu}
\end{eqnarray}
The curvature tensor is 
\begin{eqnarray}
R^{\mu}{ }^{\ }_{\nu\alpha\beta}
&=&
\partial^{\ }_{\alpha}
\Gamma^{\mu}{ }^{\ }_{\nu\beta}
-
\partial^{\ }_{\beta}
\Gamma^{\mu}{ }^{\ }_{\nu\alpha}
+
\Gamma^{\mu}{ }^{\ }_{\sigma\alpha}
\Gamma^{\sigma}{ }^{\ }_{\nu\beta}
-
\Gamma^{\mu}{ }^{\ }_{\sigma\beta}
\Gamma^{\sigma}{ }^{\ }_{\nu\alpha},
\nonumber \\%%%%%
R_{\mu\alpha}
&=&
g^{\nu\beta} R^{\ }_{\mu\nu\alpha\beta}.
\end{eqnarray}
% For the Levi-Civita connection, the Riemann curvature
% takes the explicit form
% \begin{eqnarray}
% R^{\gamma}_{\ \delta \alpha \beta }&=&
% g^{\gamma\bar \gamma}R^{\ }_{\bar \gamma \delta \alpha \beta },
% \nonumber\\
% R^{\ }_{\bar \gamma \delta \alpha \beta }&=&
% \frac{1}{2}
% \big(
% \partial^{\ }_{\alpha}
% \partial^{\ }_{\delta}
% g^{\ }_{\bar \gamma \beta }
% -
% \partial^{\ }_{\alpha}
% \partial^{\ }_{\bar \gamma}
% g^{\ }_{\beta \delta}
% -
% \partial^{\ }_{\beta }
% \partial^{\ }_{\delta}
% g^{\ }_{\bar \gamma \alpha}
% +
% \partial^{\ }_{\beta }
% \partial^{\ }_{\bar \gamma}
% g^{\ }_{\alpha \delta}
% \big)
% \nonumber\\
% &-&
% \frac{1}{4}
% \big(
% \partial^{\ }_{\alpha} g^{\ }_{\mu \bar \gamma}
% +
% \partial^{\ }_{\bar \gamma} g^{\ }_{\mu \alpha}
% -
% \partial^{\ }_{\mu} g^{\ }_{\alpha \bar \gamma}
% \big)
% g^{\mu \nu}
% \big(
% \partial^{\ }_{\beta } g^{\ }_{\nu \delta}
% +
% \partial^{\ }_{\delta} g^{\ }_{\nu \beta }
% -
% \partial^{\ }_{\nu } g^{\ }_{\beta \delta}
% \big)
% \nonumber\\
% &+&
% \frac{1}{4}
% \big(
% \partial^{\ }_{\beta } g^{\ }_{\mu \bar \gamma}
% +
% \partial^{\ }_{\bar \gamma} g^{\ }_{\mu \beta }
% -
% \partial^{\ }_{\mu} g^{\ }_{\beta \bar \gamma}
% \big)
% g^{\mu \nu}
% \big(
% \partial^{\ }_{\alpha} g^{\ }_{\nu \delta}
% +
% \partial^{\ }_{\delta} g^{\ }_{\nu \alpha}
% -
% \partial^{\ }_{\nu} g^{\ }_{\alpha\delta}
% \big).
% \label{eq: lowering upper index in curvature tensor if Levi-Civita}
% \end{eqnarray}
The curvature tensor can also be constructed from the spin connection:
\begin{eqnarray}
R^{{a}}{ }_{{b}}
=
d \omega^{{a}}{ }_{{b}}
+
\omega^{{a}}{ }_{{c}}
\wedge \omega^{{c}}{ }_{{b}}
=
R_{\mu\nu}{ }^{{c}}{ }_{{b}}dx^{\mu}dx^{\nu}
\end{eqnarray}
where
$R_{\mu\nu}{ }^{{c}}{ }_{{b}}=
R_{\mu\nu}{ }^{\rho}{ }_{\lambda}
e^c{ }_{\rho}
e_b { }^{\lambda}
$.
\bibitem{FootnoteCentralChargeEdge}
The quantity $c$ denotes the conformal central charge of the
conformal field field theory describing the (topologically protected)
chiral edge modes that would appear at a spatial (1+1) dimensional boundary.
\bibitem{WillExplainAnomalies}
We will give a brief explanation of the relevant
concepts below. See also 
Refs.\ \onlinecite{nakahara,fujikawa,AlvarezGaumeWitten83,AlvarexGaumeGinsparg85}.
\bibitem{footnote2}
It is known\cite{FosterLudwigPRB} that symmetry class
AIII can also be realized as a quasiparticle system within a (spinful) superconducting  ground state
which conserves one component (say the $S_z$-component) of Pauli-spin. In this case
the $U(1)$ charge associated with the conservation of $S_z$ can be used in lieu of
the conserved particle number of a normal (not superconducting) system in class AIII.
\bibitem{footnote3}
As explained in
the paragraph below Eq.\ (\ref{ChiralAnomalyInBackgroundGaugeField}),
the quantity ${\cal A}_{2n+2}(x)$ appearing in this
equation is given by Eqs.\ (\ref{DescentRelation}) and (\ref{DEFAnomalyPolynomial})
below, and vanishes in the absence of the electromagnetic
field strength $F_{\mu\nu}$. Therefore, this
anomaly is called ``chiral (axial) anomaly in a background $U(1)$
gauge field''.
\bibitem{footnote4}
The argument is essentially the
same as that presented in 
Subsec.\ \ref{subsec:gravitational instanton term}.
As explained below, the calculations performed in this
subsection amount to a derivation of what we will
call below a ``chiral $U(1)$ gauge anomaly in the
presence of a background gravitational field''.
\bibitem{footnote5}
The appearance of this
term for the non-Abelian gauge group $SU(2)$ was first
pointed out in the context of topological insulators (superconductors)
in Ref.\ \onlinecite{diamondCI}
%\onlinecite{RyuSchnyderLudwigClassCI} 
for the spin-singlet
topological superconductor in symmetry class CI in $d=3$ spatial dimensions.
\bibitem{footnote-Explain-BDI} 
For class BDI there exist two distinct
physical realizations, one as (``spinless'' time-reversal
invariant) superconductors, and one as normal (non-superconducting) electronic
systems. Without considering interactions
there is basically no difference between the two, except that the
number of species of Majorana fermions is even in the latter case,
where a pair of Majorana fermions is thought to be combined into
a complex fermion, carrying a $U(1)$ charge, or particle number.
The discussion of anomalies, considered in the current Section of this article,
is aimed at the discussion of interacting theories (as explained, e.g., in
the introduction). Now, when inclusion of interactions is considered,
the two above-mentioned realizations of symmetry class BDI behave
very differently. Obviously, in the latter (normal, non-superconducting)
realization, the interactions are to respect the $U(1)$ symmetry, whereas
in the former (superconducting) realization, there is no such constraint
on the form of the interactions. In this Subsection we will consider
solely the latter realization. In this case, there is thus always a conserved
$U(1)$ quantity. The former (superconducting) case was discussed recently
in Refs. \onlinecite{FidkowskiKitaev1,FidkowskiKitaev2,TurnerPollmannBerg}. 
(For similar methods applied to gapped spin chain, see, e.g.,  Ref.\ \onlinecite{Wen}.)
At present, we do not have an understanding of that case in terms of anomalies.
We hope to be able to address this case in future work.
\end{thebibliography}

%\tableofcontents
 
\end{document}